\documentstyle[12pt]{article}
\def\zid{1\kern-0.36em\llap~1}

\newcommand{\beq}{\begin{equation}}
\newcommand{\ber}{\begin{eqnarray}}
\newcommand{\eeq}{\end{equation}}
\newcommand{\eer}{\end{eqnarray}}

\begin{document}

\begin{titlepage}
\rightline{[SUNY BING 5/21/05 v.4 ] }  \rightline{ hep-ph/0506240}
\vspace{2mm}
\begin{center}
{\bf \hspace{1.85 cm} \center USE OF W-BOSON
LONGITUDINAL-TRANSVERSE INTERFERENCE \newline IN TOP QUARK SPIN-CORRELATION FUNCTIONS }\\
\vspace{2mm} Charles A.
Nelson\footnote{Electronic address: cnelson @ binghamton.edu  },
 Eric Gasparo Barbagiovanni, Jeffrey J. Berger, \newline Elisa K. Pueschel, and Joshua R. Wickman  \\
{\it Department of Physics, State University of New York at
Binghamton\\ Binghamton, N.Y. 13902}\\[2mm]
\end{center}


\begin{abstract}

Most of this paper consists of the derivation of general
beam-referenced stage-two spin-correlation functions for the
analysis of top-antitop pair-production at the Tevatron, at the
Large Hadron Collider, and/or at an International Linear Collider.
However, for the charged-lepton-plus-jets reaction $q \bar{q}
\rightarrow t \bar{t} \rightarrow (W^+ b) (W^- \bar{b} )
\rightarrow  (l^{+} \nu b ) (W^- \bar{b} )$, there is a simple
3-angle spin-correlation function for determination of the
relative sign of, or for measurement of a possible non-trivial
phase between the two dominant $\lambda_b = -1/2 $ helicity
amplitudes for the $t\rightarrow W^{+}b$ decay mode. For the
$CP$-conjugate case, there is an analogous function and tests for
$\bar{t} \rightarrow W^{-} \bar{b}$ decay. These results make use
of W-boson longitudinal-transverse interference.

\end{abstract}

\end{titlepage}

\section{ Introduction: W-Boson Longitudinal-Transverse \newline Interference }

In part because of the large top-quark mass [1] and properties of
QCD, W-boson polarimetry is a particularly powerful technique for
empirical investigation of the $t\rightarrow W^{+}b$ decay mode
from top-antitop pair-production data for the ``charged-lepton
plus jets" channel [2].  For this channel, there is the sequential
decay $t\rightarrow W^{+}b \rightarrow (l^{+} \nu) b $, with
$\bar{t} \rightarrow W^{-} \bar{b}$ in which the $W^-$ decays into
hadronic jets.  Since the final state is the $(l^{+} \nu)$ decay
product of the $W^+$, there are observable effects from $W^+$
boson longitudinal-transverse interference.  For instance, a
contribution to the angular-distribution intensity-function is the
product of an amplitude in which the $W^+$ is
longitudinally-polarized with the complex-conjugate of an
amplitude in which the $W^+$ is transversely polarized, summed
with the complex-conjugate of this product. The helicity formalism
[3] is a general method for investigating applications of W-boson
interference in stage-two spin-correlation functions for
describing the charged-lepton plus jets channel, and for the
di-lepton plus jets channel.

Most of this paper consists of the derivation of general
beam-referenced stage-two \newline spin-correlation functions
(BR-S2SC) [4-7] for the analysis of top-antitop pair-production at
the Tevatron [1], at the Large Hadron Collider [8], and/or at an
International Linear Collider [9]. However, as a simple result
which illustrates W-boson longitudinal-transverse interference,
for the charged-lepton-plus-jets reaction $q \bar{q} \rightarrow t
\bar{t} \rightarrow (W^+ b) (W^- \bar{b} ) \rightarrow  (l^{+} \nu
b ) (W^- \bar{b} )$ we have found that there is a 3-angle
spin-correlation function for (i) determination of the relative
sign of [10,11], or for (ii) measurement of a possible non-trivial
phase between the two dominant $\lambda_b = -1/2 $ helicity
amplitudes for the $t\rightarrow W^{+}b$ decay mode [12]. For the
$CP$-conjugate case, there is an analogous function and tests for
$\bar{t} \rightarrow W^{-} \bar{b}$ decay.

Tests for non-trivial phases in top-quark decays are important in
searching for possible $\widetilde T_{FS}$ violation. $\widetilde
T_{FS}$ invariance will be violated if either (i) there is a
fundamental violation of canonical time-reversal invariance,
and/or (ii) there are absorptive final-state interactions.  For
instance, unexpected final-state interactions might be associated
with additional t-quark decay modes.  To keep this assumption of
``the absence of final-state interactions" manifest in comparison
to a detailed-balance or other direct test for fundamental
time-reversal invariance, we refer to this as $\widetilde T_{FS}$
invariance, see [13,5]. Measurement of a non-zero primed top-quark
decay helicity parameter, such as $  \eta ^{\prime }$ or $ \omega
^{\prime } $, would imply $\widetilde T_{FS}$ violation, see
Appendix B. ``Explicit $\widetilde T_{FS}$ violation" will occur
[12] if there is an additional complex-coupling $ \frac {g_i}{{2
\Lambda_i}} $ associated with a specific single additional Lorentz
structure, $ i=S,P, S \pm P,...$.

For the sequential decay $ t \rightarrow W^{+} b $ followed by $
W^+ \rightarrow l^+ \nu  $, the spherical angles $ \theta_a$,
$\phi_a $ specify the $ l^+ $ momentum in the ${W_1}^+$ rest frame
(see Fig. 1) when there is first a boost from the $ (t \bar{t}
)_{c.m.}$ frame to the $t_1$ rest frame, and then a second boost
from the $t_1$ rest frame to the ${W_1}^+$ rest frame.   The $0^o$
direction for the azimuthal angle $\phi_a$ is defined by the
projection of the ${W_2}^-$ momentum direction. Correspondingly
(see Fig. 2) the spherical angles $ \theta_b$, $\phi_b $ specify
the $ l^- $ momentum in the ${W_2}^-$ rest frame when there is
first a boost from the $ (t \bar{t} )_{c.m.}$ frame to the
$\bar{t_2}$ rest frame, and then a second boost from the
$\bar{t_2}$ rest frame to the ${W_2}^-$ rest frame. The $0^o$
direction for the azimuthal angle $\phi_b$ is defined by the
projection of the ${W_1}^+$ momentum direction. As shown in Fig.
3, the two angles $\theta _1^t$, $\theta _2^t$ describe the
$W$-boson momenta directions in the first
stage of the sequential-decays of the $t\bar{t}$ system, in which $%
t_1 \rightarrow {W_1}^{+}b$ and $\bar{t_2} \rightarrow
{W_2}^{-}\bar b$.  Through out this paper, the subscripts ``one"
and ``two" will be used to distinguish the two sequential-decay
chains.

In the  $t_1$ rest frame, the matrix element for $t_1 \rightarrow
{W_1}^{+} b$ is
\begin{equation}
\langle \theta _1^t ,\phi _1 ,\lambda _{W^{+} } ,\lambda _b |\frac
12,\lambda _1\rangle =D_{\lambda _1,\mu }^{(1/2)*}(\phi _1 ,\theta
_1^t ,0)A\left( \lambda _{W^{+} } ,\lambda _b \right)
\end{equation}
where $\mu =\lambda _{W^{+} } -\lambda _b $ in terms of the
${W_1}^+$ and $b$-quark helicities. Through out this paper an
asterisk will denote complex conjugation.  The final ${W_1}^{+}$
momentum is in the $\theta _1^t ,\phi _1$ direction and the
$b$-quark momentum is in the opposite direction. The variable
$\lambda_1$ gives the $t_1$-quark's spin component quantized along
the $z_{1}^t$ axis in Fig. 3.  Upon a boost back to the $(t\bar
t)_{cm}$, or on further to the $\bar{t_2}$ rest frame, $\lambda_1$
also specifies the helicity of the $t_1$-quark. For the
$CP$-conjugate process, $\bar t_2 \rightarrow {W_2}^{-} \bar b$,
in the $\bar t_2$ rest frame the matrix element is
\begin{equation}
\langle \theta _2^t ,\phi _2 ,\lambda _{W^{-} },\lambda _{\bar
b}|\frac 12,\lambda _2\rangle =D_{\lambda _2,\bar \mu
}^{(1/2)*}(\phi _2 ,\theta _2^t ,0)B\left( \lambda _{W^{-}
},\lambda _{\bar b}\right)
\end{equation}
with $\bar \mu =\lambda _{W^{-}}-\lambda _{\bar b}$. By analogous
argument, $\lambda_2$ is the $\bar t_2$ helicity.

In terms of the $t\rightarrow W^{+}b$ helicity amplitudes, the
polarized-partial-widths and W-boson-LT-interference-widths are
\begin{eqnarray}
\Gamma (0,0) & \equiv &\left| A(0,-1/2)\right| ^{2}, \, \, \Gamma
(-1,-1)\equiv
\left| A(-1,-1/2)\right| ^{2}   \\
\Gamma _{\mathit{R}}(0,-1) &=&\Gamma _{\mathit{R}}(-1,0)\equiv {{
Re}[A(0,-1/2)A(-1,-1/2)^{\ast }}]  \nonumber \\
& \equiv & |A(0,-1/2)||A(-1,-1/2)|\cos \beta _{L}  \\
\Gamma _{\mathit{I}}(0,-1) &=&-\Gamma _{\mathit{I}}(-1,0) \equiv {Im}%
[A(0,-1/2)A(-1,-1/2)^{\ast }]  \nonumber \\
& \equiv &-|A(0,-1/2)||A(-1,-1/2)|\sin \beta _{L}
\end{eqnarray}
where the $R$, $I$ subscripts denote the real and imaginary parts
which define the W-boson-LT-interference. The ${L}$ superscript on the $%
\Gamma ^{{L}}(\lambda _{W},\lambda _{W}^{^{\prime }})$'s has been
conveniently suppressed in (3-5) for this is the dominant
$\lambda_b$ helicity channel. \ By convention, the dominant ${L}$
superscript [ ${R}$ superscript ] on $ \Gamma ^{{L}}(\lambda
_{W},\lambda _{W}^{^{\prime }})$ [  $ \overline{{\Gamma
}}^{R}(\lambda _{W},\lambda _{W}^{^{\prime }})$ ]  will be
suppressed in this paper. Note the two important minus-signs in
the last two lines of (5). Here, following the conventions in
[5,11,12,14], we define the moduli and phases as
\begin{equation}
A(\lambda _{W},\lambda _{b})\equiv \left| A(\lambda _{W},\lambda
_{b})\right| \exp (\imath \, \, \varphi _{\lambda _{W},\lambda
_{b} })
\end{equation}
with
\begin{equation}
\beta _{L}  \equiv  \varphi _{-1,-\frac{1}{2}}-\varphi
_{0,-\frac{1}{2}}, \, \,  \beta _{R}  \equiv  \varphi
_{1,\frac{1}{2}}-\varphi _{0,\frac{1}{2}}
\end{equation}

In terms of the $\overline{t}\rightarrow W^{-}\overline{b}$
helicity amplitudes,
\begin{eqnarray}
\overline{\Gamma }(0,0) & \equiv &\left| B(0,1/2)\right| ^{2}, \, \, \overline{%
\Gamma }(1,1)\equiv \left| B(1,1/2)\right| ^{2}  \\
\overline{\Gamma }_{R}(0,1) &=&\overline{\Gamma }_{\mathit{R}%
}(1,0)\equiv {{Re}[B(0,1/2)B(1,1/2)^{\ast }}]  \nonumber \\
& \equiv &|B(0,1/2)||B(1,1/2)|\cos \overline{\beta }_{R}  \\
\overline{\Gamma }_{I}(0,1) &=& - \overline{\Gamma }_{I}(1,0)
 \equiv {{Im} [B(0,1/2)B(1,1/2)^{\ast }}]  \nonumber \\
& \equiv &-|B(0,1/2)||B(1,1/2)|\sin \overline{\beta }_{R}
\end{eqnarray}
with the moduli and phases defined by
\begin{equation} B(\lambda
_{W},\lambda _{\overline{b}})\equiv \left| B(\lambda _{W},\lambda
\overline{_{b}})\right| \exp (\imath \, \, \overline{\varphi
}_{\lambda _{W},\lambda \overline{_{b}}})
\end{equation}
with $ \overline{\beta }_{R}  \equiv \overline{\varphi }_{1,\frac{1}{2}}-\overline{%
\varphi }_{0,\frac{1}{2}}$ and  $ \overline{\beta }_{L} \equiv  \overline{\varphi }_{-1,-\frac{1}{2}}-%
\overline{\varphi }_{0,-\frac{1}{2}} $.

In this paper, we consider the production-decay sequence
\begin{equation}
q \overline{q} , \, {\mathrm{ or }} \,  \, e\bar{e}\rightarrow
t\overline{t}\rightarrow (W^{+}b)(W^{-}%
\overline{b}) \rightarrow \cdots
\end{equation}
At the Tevatron, this is the dominant contribution to $t \bar{t}$
production.  The contribution from $gg \rightarrow
t\overline{t}\rightarrow (W^{+}b)(W^{-}%
\overline{b}) \rightarrow \cdots$ can be treated analogously.  The
latter is the dominant contribution at the LHC. The corresponding
BR-S2SC functions for it will be reported separately [15].

We assume that the $\lambda _{b}=-1/2$ and $\lambda \overline{_{b}}$ $%
=1/2$ amplitudes dominate respectively in $t_1$ and $\bar{t_2}$
decay. \ In the SM and in the case of an additional large $t_R
\rightarrow b_L $ moment [10], the $\lambda _{b}=-1/2$ and $%
\lambda \overline{_{b}}$ $=1/2$ amplitudes are more than $\sim 30$
times larger than the $\lambda _{b}=1/2$ and $\lambda
\overline{_{b}}$ $=-1/2$ amplitudes. The simple three-angle
distribution ${ \mathcal{F}{|}} _{0} +
 {\mathcal{F}{|}}_{sig} $ for $%
{t}_{1}\rightarrow W_{1}^{+}{b}\rightarrow ( l^{+}\nu ) {b}$
involves the angles $\{ \theta _{2}^{t}$, $ \theta _{a}$, $\phi
_{a} \}$ shown in Figs. 1-3.
\begin{equation}
{ \mathcal{F}{|}} _{0} {\mathcal{=}}\frac{16\pi ^{3}g^{4}}{9s^{2}}(1+\frac{2m_{t}^{2}%
}{s})\left\{ \frac{1}{2}\Gamma (0,0)\sin ^{2}\theta _{a}+\Gamma
(-1,-1)\sin ^{4}\frac{\theta _{a}}{2}\right\} [\overline{\Gamma }(0,0)+%
\overline{\Gamma }(1,1)]
\end{equation}
\begin{eqnarray}
{\mathcal{F}{|}}_{sig} &=&{-}\frac{4\sqrt{2}\pi ^{4}g^{4}}{9s^{2}}(1+%
\frac{2m_{t}^{2}}{s})\cos \theta _{2}^{t} \sin \theta _{a}\sin ^{2}\frac{%
\theta _{a}}{2} [\overline{\Gamma }(0,0)+%
\overline{\Gamma }(1,1)] \nonumber \\
&&\left\{ \Gamma _{R}(0,-1)\cos \phi _{a}-\Gamma _{I}(0,-1)\sin
\phi _{a}\right\} K {\mathcal{R}}
\end{eqnarray}
where $K$, ${\mathcal{R}}$ are defined below.

The analogous three-angle S2SC function $ \overline{{ \mathcal{F}} {|}} _{0}
+ \overline{{\mathcal{F}} {|}}_{sig} $ for the $CP$-conjugate channel $%
\overline{t}_{2}\rightarrow W_{2}^{-}\overline{b}\rightarrow (
l^{-}\bar{\nu} ) \overline{b}$ is a distribution versus  $\{
\theta _{1}^{t}$, $ \theta _{b}$, $\phi _{b} \}$ :
\begin{eqnarray}
\overline{{ \mathcal{F}} {|}} _{0} {\mathcal{=}}\frac{16\pi
^{3}g^{4}}{9s^{2}}
 (1+\frac{2m_{t}^{2}}{s})\left\{
\frac{1}{2}\overline{\Gamma }(0,0)\sin
^{2}\theta _{b}+\overline{\Gamma }(1,1)\sin ^{4}\frac{\theta _{b}}{2}%
\right\} [\Gamma (0,0)+\Gamma (-1,-1)]
\end{eqnarray}
\begin{eqnarray}
\overline{{\mathcal{F}} {|}}_{sig} &=&{-}\frac{4\sqrt{2}\pi
^{4}g^{4}}{9s^{2}} (1+\frac{2m_{t}^{2}}{s})\cos \theta _{1}^{t}
\sin
\theta _{b}\sin ^{2}\frac{\theta _{b}}{2} [\Gamma (0,0)+\Gamma (-1,-1)] \nonumber \\
&&\left\{ \overline{\Gamma }_{R}(0,1)\cos \phi _{b}+\overline{\Gamma }%
_{I}(0,1)\sin \phi _{b}\right\} K \overline{{\mathcal{R}}}
\end{eqnarray}
Note the important relative plus-sign between $\overline{\Gamma
}_{I}(0,1)$ and $\overline{\Gamma }_{R}(0,1)$ in (16), in contrast
to the relative minus-sign for ${\Gamma }_{I}(0,1)$ and ${\Gamma
}_{R}(0,1)$ in (14).

\subsection{Structure of three-angle S2SC functions}

The ``signal" contributions are suppressed by the factor
\begin{equation} K\equiv
\frac{(1-\frac{2m_{t}^{2}}{s})}{(1+\frac{2m_{t}^{2}}{s})}
\end{equation}
associated with the $g \rightarrow t \bar{t} $ production process,
and the factor
\begin{equation} {\mathcal{R}} \equiv \frac{[{\overline{\Gamma
}}(0,0)-{\overline{\Gamma }}(1,1)]} {[{\overline{\Gamma
}}(0,0)+{\overline{\Gamma }}(1,1)]}, \, \overline{{\mathcal{R}}}
\equiv  \frac{[{\Gamma }(0,0)-{\Gamma }(-1,-1)]} {[{\Gamma
}(0,0)+{\Gamma }(-1,-1)]}
\end{equation}
associated with the stage-one part of the sequential-decay chains,
$\bar{t}  \rightarrow W^{-} \bar{b}, {t}  \rightarrow W^{+} {b}$.
Numerically, ${\mathcal{R}} \sim 0.41$ in both the standard model
and in the case of an additional large $t_R \rightarrow b_L$
chiral weak-transition moment [10]. The appearance of the
${\mathcal{R}} = ({\mathtt{prob}} \, W_L) - ({\mathtt{prob}} \,
W_T) $ factor is not surprising [4,13] because this is a
consequence of the dynamical assumption that the
$\lambda _{b}=-1/2$ and $\lambda \overline{_{b}}$ $%
=1/2$ amplitudes dominate. In the standard model ${\mathcal{R}} =
(1-\frac{2m_{W}^{2}}{{m_t}^{2}}) /
(1+\frac{2m_{W}^{2}}{{m_t}^{2}})$ whether there is or isn't a
large $t_R \rightarrow b_L$ moment. Fortunately $m_t \neq \sqrt{2}
m_W =+ 113GeV$, otherwise many $W$-boson polarimetry effects would
be absent in top-quark spin-correlation functions. An important
exception is the $\theta_a$ dependence of ${ \mathcal{F}{|}} _{0}$
[ see (13)].  Both of the ${\mathcal{R}}$ and $K$ suppression
factors are absent in purely stage-two $W$-boson polarimetry, with
or without spin-correlation.

From the ${\theta_2}^t$ dependence of the integrated
diagonal-elements of the sequential-decay density matrices for
$\bar{t_2} \rightarrow {W_2}^- \bar{b} \rightarrow ( l^- \bar{\nu}
) \bar{b} $, it follows that ${\mathcal{R}}$'s numerator appears
in ${\mathcal{F}{|}}_{sig}$ multiplied by $\cos \theta _{2}^{t}$
and that ${\mathcal{R}}$'s denominator appears in
${\mathcal{F}{|}}_{0}$ multiplied by one [ see (95-96)]. Because
the t-quark has spin $\frac{1}{2}$, there are purely half-angle
$d_{m m^{\prime}} ^{\frac{1}{2} } ( {\theta_2}^t ) $-squared
intensity-product-factors in (95-97). The off-diagonal
$\overline{R} _{\lambda _2\lambda _2^{^{\prime }}}$ elements which
describe $\bar{t_2}$-helicity interference do not contribute due
to the integration over the opening-angle $\phi$ between the $t_1$
and $\bar{t_2}$ decay planes. The angles $ \theta_{1,2} $ are
respectively equivalent to the ${W_{1,2}}^{\pm}$-boson energies in
the $(t\bar t)_{cm}$ (see Appendix A).  In this 3-variable
spin-correlation function, the minus sign in the numerator of the
$K$ suppression factor in ${\mathcal{F}{|}}_{sig}$ is a
consequence of the minus sign in the sequential-decay
density-matrix $\mathbf{R}_{++}^{b_{L}}$ of (26) in the
helicity-flip contribution (92) for the
$\overline{\mathbf{R}}_{++}$ term, versus the corresponding plus
sign in $\mathbf{R}_{--}^{b_{L}} $ of (27) in the
helicity-conserving contribution (72) for the
$\overline{\mathbf{R}}_{++}$ term; and analogously for the
$\overline{\mathbf{R}}_{--}$ terms in (92) and (72).

\subsection{Summary}

From the top-quark spin-correlation function (13-14), the two
tests for $t_1 \rightarrow {W_1}^+ b $ decay are:

(i) By measurement of ${\Gamma }_{R}(0,-1)$, the relative sign of
the two dominant $\lambda_b=-1/2$ helicity-amplitudes can be
determined if their relative phase is $0^0$ or $180^0$. Versus the
partial-decay-width $\Gamma ( t \rightarrow W^+ b ) $, W-boson
longitudinal-transverse interference is a large effect for in the
standard model $\eta_L \equiv \frac{{\Gamma }_{R}(0,-1)}{\Gamma} =
\pm 0.46 $ without/with a large $ t_R \rightarrow b_L $ chiral
weak-transition-moment.  In both models, the probabilities for
longitudinal/transverse W-bosons are large, $P(W_{L}) =
\frac{{\Gamma }(0,0)}{\Gamma} = 0.70$ and $P(W_{T}) =
\frac{{\Gamma }(-1,-1)}{\Gamma} = 0.30$, and so for a trivial
relative-phase difference of $0^0$ or $180^0$, W-boson
longitudinal-transverse interference must be a large effect.

(ii) By measurement of both ${\Gamma }_{R}(0,-1)$ and ${\Gamma
}_{I}(0,-1)$ via the $\phi_a$ dependence, a possible non-trivial
phase can be investigated.  Tests for non-trivial phases in
top-quark decays are important in searching for possible
$\widetilde T_{FS}$ violation.

From (15-16), there are the analogous two tests for $\bar{t}_2
\rightarrow {W_2}^- \bar{b} $ decay.  In the standard model $
\overline{\Gamma }_{R}(0,1) = {\Gamma }_{R}(0,-1)$, and both
$\overline{\Gamma }_{I}(0,1)$ and $ {\Gamma }_{I}(0,-1)$ vanish
whether there is or isn't a purely-real $ t_R \rightarrow b_L $
transition-moment.

Section 2 of this paper contains the derivation of general BR-S2SC
functions.  For $t \bar{t}$ production by $q \bar{q}$, or $e
\bar{e} \rightarrow t \bar{t}$, neither $CP$ invariance nor
$\widetilde T_{FS}$ invariance is assumed for the $T( \lambda_1,
\lambda_2 )$ helicity amplitudes in Sec. 2.2. For informative
details, see [16].   By $CP$ invariance, $T(++)=T(--)$ but $T(+-)$
and $T(-+)$ are unrelated.  If experiment were to show that one of
the primed production-helicity-parameters (76, 82-85, 94) is
non-zero, then $\widetilde T_{FS}$ invariance is violated in the $
g \rightarrow t \bar{t} $ process.

In Section 3, these results are applied to the lepton-plus-jets
channel of the $t \bar{t}$ system, assuming that the $\lambda_b =
-1/2$ and $ \lambda_{\bar{b}} =1/2$ amplitudes dominate.  Simple
four-angle spin-correlation functions are obtained, which do not
involve beam-referencing.  These and other additional-angle
generalizations might be useful empirically, for instance as
checks with respect to the above four tests. Section 4 contains a
discussion. The appendices respectively treat (A) kinematic
formulas,  (B) translation between this paper's $\Gamma (
\lambda_W , {\lambda_W}^{'} )$ notation and the helicity
parameter's notation of Refs. [5,11,12,14], (C) kinematic formulas
for beam-referencing versus Figs. 1-2, and (D) formulas for $e
\bar{e} \rightarrow t \bar{t}$ production.

\section{Derivation of Beam-Referenced Stage-Two \newline
 Spin-Correlation Functions}

In order to reference stage-two spin-correlation functions (S2SC)
to the incident lepton or parton beam [4], we generalize the
derivation of S2SC functions given in [5]. When more data is
available for top quark decays, it should be a reasonable further
step to consider using the results of [14] to incorporate
$\Lambda_b$ polarimetry.   $\Lambda_b$ polarimetry could be used
to make a complete measurement of the four moduli and the three
relative-phases of the helicity amplitudes in $t \rightarrow
{W}^{+}b$ and analogously in $\bar{t} \rightarrow W^- \bar{b}$. In
this context, next-to-leading-order QCD, electroweak, and W-boson
and t-quark finite-width corrections require further theoretical
investigation [7].  If the magnitudes of the two $\lambda_b = 1/2$
helicity amplitudes are as predicted by the standard model, i.e.
at factors of more than $ \sim \frac{1}{30}$ smaller than the two
dominant $\lambda_b = -1/2$ amplitudes, both detector and
background effects will be non-trivial at this level of
sensitivity at a hadron collider. Nevertheless, empirical
consideration will be warranted if by then, there is compelling
evidence for unusual top-quark physics.

In the BR-S2SC functions, we consider the decay sequence $t_1
\rightarrow {W_1}^{+}b$ followed by
 ${W_1}^{+} \rightarrow l^{+} \nu$, and the $CP$-conjugate decay sequence $\bar{t_2}
\rightarrow {W_2}^{-} \bar{b}$ followed by
 ${W_2}^{-} \rightarrow l^{-} \bar{\nu}$.  In Figs. 3 and 4, the spherical angles
$\theta _1^t$ and $ \phi _1$  describe the ${W_1}^+$ momentum in
the ``first stage" $t_1 \rightarrow {W_1}^{+}b$. Similarly, in
Fig. 5 spherical angles $\theta _a$ and $ \tilde{\phi _a}$
describe the ${l}^+$ momentum in the ``second stage" ${W_1}^+
\rightarrow {l}^{+} \nu$ when there is first a boost from the
$(t\bar t)_{cm}$ frame to the $t_1$ rest frame, and then a second
boost from the $t_1$ rest frame to the ${W_1}^+$ rest frame. If
instead the boost to the ${W_1}^{+}$ rest frame is directly from
the $(t\bar t)_{cm}$ frame, one must account for Wigner rotations.
Formulas and details about these Wigner rotations are given in
Ref. [5]. Analogously, two pairs of spherical angles
${\theta_2}^t, \phi_2$ and $\theta_b$, $ \tilde{\phi_b}$ specify
the two stages in the $CP$-conjugate sequential decay $ \bar{t}
\rightarrow W^{-} \bar{b} $ followed by $ W^- \rightarrow l^-
\bar{ \nu} $ when the boost is from the $\bar{t_2}$ rest frame.

Note that the charged leptons' azimuthal angle $ \tilde{\phi _a}$
in the ${W_1 }^+$ rest frame in Fig. 5, and analogously $
\tilde{\phi_b}$ in the ${W_2}^-$ rest frame, are referenced
respectively by the $\bar{t_2}$ and $ t_1$ momentum directions.
Instead of using the anti-top and top quark momenta for this
purpose, one can reference these two azimuthal angles in terms of
the opposite $W^{\mp}$-boson momentum as in the formulas given in
the introduction. These azimuthal angles are then denoted without
``tilde accents" : $ \phi _a $ in the ${W_1 }^+$ rest frame when
the boost is from the $t_1$ rest frame, and $ \phi_b $ in the
${W_2}^-$ rest frame when the boost is from the $\bar{t_2}$ rest
frame.

As discussed in the caption to Fig. 3, the momenta for $t_1$,
${W_1 }^+$, and $\bar{t_2}$ lie in the same plane whether the
analysis is in the $t_1$ rest frame, in the $\bar{t_2}$ rest
frame, or in the  $t\bar{t}$ center-of-momentum frame.  Therefore,
in deriving BR-S2SC functions in the helicity formalism, the angle
$ \tilde{\phi _a}$ in the ${W_1 }^+$ rest frame is theoretically
clear and simple. In general in the $(t\bar t)_{cm}$ frame, the
momenta for $t_1$, ${W_1 }^+$ and ${W_2 }^-$ do not lie in the
same plane. However, from the empirical point of view, the ${W_2
}^-$ momentum direction in the ${W_1 }^+$ rest frame will often be
more precisely known, and so these two azimuthal angles without
``tilde accents" will be more useful. From the standpoint of the
helicity formalism, in the final S2SC functions either $\phi_a$ or
$\tilde{\phi_a}$ can be used because it is only a matter of
referencing the zero direction for the azimuthal angle, i.e. it is
an issue concerning the specification of the Euler angles in the
$D$ function for $ W^+ \rightarrow l^+ \nu  $ decay.

To simplify the notation, unlike in Refs. [5,14], in this paper we
do not use ``tilde accents" on the polar angles $\theta_a$ and
$\theta_b$.  We also do not use ``$t$" superscripts on
$\phi_{1,2}$ for they are Lorentz invariant for each of the three
frames considered in Fig. 3.  On the other hand, ``$t$"
superscripts on ${\theta_{1,2} }^t$ for the $t_1$ and $\bar{t_2}$
rest frames, are necessary to distinguish these angles from
$\theta_{1,2}$ which are defined in the $(t\bar t)_{cm}$.

In the ${W_1}^{+}$ rest frame, the matrix element for ${W_1}^{+}
\rightarrow l^+ \nu$ [ or for ${W_1}^{+} \rightarrow j_{\bar d}
j_{u}$ ] is
\begin{equation}
\langle \theta _a ,\tilde \phi _a ,\lambda _{l^+} ,\lambda _{\nu}
| 1,\lambda _{W^+} \rangle =D_{\lambda _{W^+},1 }^{1*}
 (\tilde \phi
_a , \theta _a ,0)c
\end{equation}
since $\lambda _{\nu}= - \frac{1}{2}, \lambda _{l^+}=
\frac{1}{2}$, neglecting $(\frac{m_l}{m_W})$ corrections [
neglecting $(\frac{ m_{jet} }{m_W})$ corrections].  Since the
amplitude ``$c$" in this matrix element is independent of the
helicities, we will suppress it in the following formulas since it
only affects the overall normalization. We will use below
\begin{equation}
\rho _{\lambda _1\lambda _1^{^{\prime }};\lambda _W\lambda
_W^{^{\prime }}}(t\rightarrow W^{+}b)=\sum_{\lambda _b=\mp
1/2}D_{\lambda _1,\mu }^{(1/2)*}(\phi _1,\theta _1^t,0)D_{\lambda
_1^{^{\prime }},\mu ^{^{\prime }}}^{(1/2)}(\phi _1,\theta
_1^t,0)A(\lambda _W,\lambda _b){A^{*}}(\lambda _W^{^{\prime
}},\lambda _b)
\end{equation}
where $\mu =\lambda _{W^{+}}-\lambda _{b}$ and $\mu ^{^{\prime
}}=\lambda _{W^{+}}-\lambda _{b}^{^{\prime }}$,
\begin{equation}
{ \rho} _{\lambda _W\lambda _W^{^{\prime }}} (W^{+}\rightarrow
l^{+}\nu )=D_{\lambda _W,1}^{1*}(\widetilde{\phi _a},{\theta
_a}%
,0)D_{\lambda _W^{^{\prime }},1}^1(\widetilde{\phi
_a},{\theta _a}%
,0)
\end{equation}
 In the ${W_2}^{-}$
rest frame, analogous to (19) the matrix element for ${W_2}^{-}
\rightarrow l^- \bar{\nu}$ [ ${W_2}^{-} \rightarrow j_{\bar u}
j_{d}$ ] is
\begin{equation}
\langle \theta _b ,\tilde \phi _b ,\lambda _{l^-} ,\lambda
_{\bar{\nu}} | 1,\lambda _{W^-} \rangle =D_{\lambda _{W^-},-1
}^{1*}
 (\tilde \phi
_b , \theta _b ,0)\bar{c}
\end{equation}
and we suppress the ``$\bar{c}$" factor in the following.

\subsection{ Sequential-decay density matrices}

The composite decay-density-matrix for $t_1 \rightarrow
{W_1}^{+}b\rightarrow (l^{+}\nu )b$ is
\begin{equation}
{{{R_{\lambda _1\lambda _1^{^{\prime }}}}} =\sum_{\lambda
_W,\lambda _W^{^{\prime }}}\rho _{\lambda _1\lambda _1^{^{\prime
}};\lambda _W\lambda _W^{^{\prime }}}(t\rightarrow W^{+}b)\rho
_{\lambda _W\lambda _W^{^{\prime }}}(W^{+}\rightarrow l^{+}\nu)}
\end{equation}
where $\lambda _W,\lambda _W^{^{\prime }}=0, \pm 1 $ and the
$\rho$ density matrices are given in (20-21).

The above composite decay-density-matrix (23)  can be expressed
\ber {\bf R=} {\bf R^ {b_L}}+ {\bf R^{b_R} } \eer

The $\lambda_b = -1/2$ elements  are
 \ber {\bf R^ {b_L}
=}\left(
\begin{array}{cc}
{\bf R^{b_L}}_{++} & e^{\imath \phi _1 } {\bf r^{b_L}}_{+-} \\
e^{-\imath \phi _1 }{\bf r^{b_L}}_{-+} & {\bf R^{b_L}}_{--}
\end{array}
\right) \eer where
\begin{eqnarray}
\mathbf{R}_{++}^{b_{L}} &=&\frac{1}{2}\Gamma (0,0)\cos
^{2}\frac{\theta
_{1}^{t}}{2}\sin ^{2}{\theta _{a}}+\Gamma (-1,-1)\sin ^{2}\frac{%
\theta _{1}^{t}}{2}\sin ^{4}{\frac{\theta _{a}}{2}}  \nonumber \\
&&-\frac{1}{\sqrt{2}}[\Gamma _{\mathit{R}}(0,-1)\cos \widetilde{\varphi _{a}}%
-\Gamma _{\mathit{I}}(0,-1)\sin \widetilde{\varphi _{a}}%
]\sin \theta _{1}^{t}\sin {\theta _{a}}\sin ^{2}{\frac{%
\theta _{a}}{2}}
\end{eqnarray}
\begin{eqnarray}
\mathbf{R}_{--}^{b_{L}} &=&\frac{1}{2}\Gamma (0,0)\sin
^{2}\frac{\theta
_{1}^{t}}{2}\sin ^{2}{\theta _{a}}+\Gamma (-1,-1)\cos ^{2}\frac{%
\theta _{1}^{t}}{2}\sin ^{4}{\frac{\theta _{a}}{2}}  \nonumber \\
&&+\frac{1}{\sqrt{2}}[\Gamma _{\mathit{R}}(0,-1)\cos \widetilde{\varphi _{a}}%
-\Gamma _{\mathit{I}}(0,-1)\sin \widetilde{\varphi _{a}}%
]\sin \theta _{1}^{t}\sin {\theta _{a}}\sin ^{2}{\frac{%
\theta _{a}}{2}}
\end{eqnarray}
\begin{eqnarray}
\mathit{{Re}}(\mathbf{r}_{+-}^{b_{L}}) &=&\frac{1}{4}\Gamma
(0,0)\sin \theta _{1}^{t}\sin ^{2}{\theta _{a}}-\frac{1}{2}\Gamma
(-1,-1)\sin
\theta _{1}^{t}\sin ^{4}{\frac{\theta _{a}}{2}}  \nonumber \\
&&+\frac{1}{\sqrt{2}}[\Gamma _{\mathit{R}}(0,-1)\cos \widetilde{\varphi _{a}}%
-\Gamma _{\mathit{I}}(0,-1)\sin \widetilde{\varphi _{a}}%
]\cos \theta _{1}^{t}\sin {\theta _{a}}\sin ^{2}{\frac{%
\theta _{a}}{2}}
\end{eqnarray}
\begin{equation}
{Im}( { \mathbf{{r}_{+-}^{b_{L}}} } ) = \frac{1}{\sqrt{2}}[\Gamma
_{\mathit{R}}(0,-1)\sin \widetilde{\varphi _{a}}+\Gamma
_{\mathit{I}}(0,-1)\cos
\widetilde{\varphi _{a}}]\sin {\theta _{a}}\sin ^{2}{%
\frac{\theta _{a}}{2} }
\end{equation}
and $\mathbf{r}_{+-}^{b_{L}}=(\mathbf{r}_{-+}^{b_{L}})^{\ast }$.

For the subdominant  \textbf{b}$_{R}$ decay channel,
\begin{equation}
\mathbf{R^{b_{R}}=}\left(
\begin{array}{cc}
\mathbf{R^{b_{R}}}_{++} & e^{\imath \phi _{1}}\mathbf{r^{b_{R}}}_{+-} \\
e^{-\imath \phi _{1}}\mathbf{r^{b_{R}}}_{-+} &
\mathbf{R^{b_{R}}}_{--}
\end{array}
\right)
\end{equation}
\begin{eqnarray}
\mathbf{R}_{++}^{b_{R}} &=&\frac{1}{2}\Gamma ^{R}(0,0)\sin
^{2}\frac{\theta
_{1}^{t}}{2}\sin ^{2}{\theta _{a}}+\Gamma ^{R}(1,1)\cos ^{2}\frac{%
\theta _{1}^{t}}{2}\cos ^{4}{\frac{\theta _{a}}{2}}  \nonumber \\
&&-\frac{1}{\sqrt{2}}[\Gamma _{\mathit{R}}^{R}(0,1)\cos
\widetilde{\varphi _{a}}+\Gamma _{\mathit{I}}^{R}(0,1)\sin
\widetilde{\varphi
_{a}}]\sin \theta _{1}^{t}\sin {\theta _{a}}\cos ^{2}{%
\frac{\theta _{a}}{2}}
\end{eqnarray}
\begin{eqnarray}
\mathbf{R}_{--}^{b_{R}} &=&\frac{1}{2}\Gamma ^{R}(0,0)\cos
^{2}\frac{\theta
_{1}^{t}}{2}\sin ^{2}{\theta _{a}}+\Gamma ^{R}(1,1)\sin ^{2}\frac{%
\theta _{1}^{t}}{2}\cos ^{4}{\frac{\theta _{a}}{2}}  \nonumber \\
&&+\frac{1}{\sqrt{2}}[\Gamma _{\mathit{R}}^{R}(0,1)\cos
\widetilde{\varphi _{a}}+\Gamma _{\mathit{I}}^{R}(0,1)\sin
\widetilde{\varphi
_{a}}]\sin \theta _{1}^{t}\sin {\theta _{a}}\cos ^{2}{%
\frac{\theta _{a}}{2}}
\end{eqnarray}
\begin{eqnarray}
\mathit{{Re}}(\mathbf{r}_{+-}^{b_{R}}) &=&-\frac{1}{4}\Gamma
^{R}(0,0)\sin \theta _{1}^{t}\sin ^{2}{\theta _{a}}+\frac{1}{2}
\Gamma ^{R}(1,1)\sin \theta _{1}^{t}\cos ^{4}{\frac{\theta
_{a}}{2}}
\nonumber \\
&&+\frac{1}{\sqrt{2}}[\Gamma _{\mathit{R}}^{R}(0,1)\cos
\widetilde{\varphi _{a}}+\Gamma _{\mathit{I}}^{R}(0,1)\sin
\widetilde{\varphi
_{a}}]\cos \theta _{1}^{t}\sin {\theta _{a}}\cos ^{2}{%
\frac{\theta _{a}}{2}}
\end{eqnarray}
\begin{equation}
{Im}({ \mathbf{r}_{+-}^{b_{R}} })=\frac{1}{\sqrt{2}}[\Gamma _{\mathit{R}%
}^{R}(0,1)\sin \widetilde{\varphi _{a}}-\Gamma _{\mathit{I}%
}^{R}(0,1)\cos \widetilde{\varphi _{a}}]\sin {\theta _{a}}\cos ^{2}%
{\frac{\theta _{a}}{2}}
\end{equation}
and $\mathbf{r}_{+-}^{b_{R}}=(\mathbf{r}_{-+}^{b_{R}})^{\ast }$ .
The \textbf{b}$_{R}$ decay channel's polarized-partial-widths and
\newline W-boson-LT-interference-widths are
\begin{eqnarray}
\Gamma ^{R}(0,0) & \equiv &\left| A(0,1/2)\right| ^{2}, \, \,
\Gamma ^{R}(1,1)\equiv
\left| A(1,1/2)\right| ^{2}   \\
\Gamma _{\mathit{R}}^{R}(0,1) &=&\Gamma
_{\mathit{R}}^{R}(1,0)\equiv
{Re}[A(0,1/2)A(1,1/2)^{\ast }]  \nonumber \\
& \equiv &|A(0,1/2)||A(1,1/2)|\cos \beta _{R}   \\
\Gamma _{\mathit{I}}^{R}(0,1) &=&-\Gamma _{\mathit{I}}^{R}(1,0)\equiv {%
Im}[A(0,1/2)A(1,1/2)^{\ast }]  \nonumber \\
& \equiv &-|A(0,1/2)||A(1,1/2)|\sin \beta _{R}
\end{eqnarray}
Note that the superscripts on these $\Gamma( \lambda_W,
{\lambda_W}^{\prime})$'s always denote the $b$ or $\bar{b}$
helicity, whereas the subscripts denote the real or imaginary part
(e.g. alternatively for (36) use $\Gamma
_{\mathit{Re}}^{R}(0,1)$).

The analogous composite decay-density matrix for the
$CP$-conjugate process \newline $ \bar{t} \rightarrow W^- \bar{b}
\rightarrow ( l^- \bar{\nu} ) \bar{b} $ is \ber {\bf \bar{R}=}
{\bf \bar{R}^ {\bar{b}_L}}+ {\bf \bar{R}^{\bar{b}_R} } \eer where
the dominant
\begin{equation}
\mathbf{\bar{R}^{\bar{b}_{R}}=}\left(
\begin{array}{cc}
\mathbf{\bar{R}^{\bar{b}_{R}}}_{++} & e^{\imath \phi _{2}}\mathbf{\bar{r}^{\bar{b}_{R}}}_{+-} \\
e^{-\imath \phi _{2}}\mathbf{\bar{r}^{\bar{b}_{R}}}_{-+} &
\mathbf{\bar{R}^{\bar{b}_{R}}}_{--}
\end{array}
\right)
\end{equation}
\begin{eqnarray}
\overline{\mathbf{R}}_{++}^{\overline{b}_{R}}
&=&\frac{1}{2}\overline{\Gamma
}(0,0)\sin ^{2}\frac{\theta _{2}^{t}}{2}\sin ^{2}{\theta _{b}}+%
\overline{\Gamma }(1,1)\cos ^{2}\frac{\theta _{2}^{t}}{2}\sin ^{4}{%
\frac{\theta _{b}}{2}}  \nonumber \\
&&+\frac{1}{\sqrt{2}}[\overline{\Gamma }_{\mathit{R}}(0,1)\cos \widetilde{%
\varphi _{b}}+\overline{\Gamma }_{\mathit{I}}(0,1)\sin
\widetilde{\varphi _{b}}]\sin \theta _{2}^{t}\sin {\theta _{b}}%
\sin ^{2}{\frac{\theta _{b}}{2}}
\end{eqnarray}
\begin{eqnarray}
\overline{\mathbf{R}}_{--}^{\overline{b}_{R}}
&=&\frac{1}{2}\overline{\Gamma
}(0,0)\cos ^{2}\frac{\theta _{2}^{t}}{2}\sin ^{2}{\theta _{b}}+%
\overline{\Gamma }(1,1)\sin ^{2}\frac{\theta _{2}^{t}}{2}\sin ^{4}{%
\frac{\theta _{b}}{2}}  \nonumber \\
&&-\frac{1}{\sqrt{2}}[\overline{\Gamma }_{\mathit{R}}(0,1)\cos \widetilde{%
\varphi _{b}}+\overline{\Gamma }_{\mathit{I}}(0,1)\sin
\widetilde{\varphi _{b}}]\sin \theta _{2}^{t}\sin {\theta _{b}}%
\sin ^{2}{\frac{\theta _{b}}{2}}
\end{eqnarray}
\begin{eqnarray}
\mathit{{Re}}(\mathbf{\overline{r}_{+-}^{\overline{b}_{R}}}) &=&-\frac{1}{4}%
\overline{\Gamma }(0,0)\sin \theta _{2}^{t}\sin ^{2}{\theta _{b}}+%
\frac{1}{2}\overline{\Gamma }(1,1)\sin \theta _{2}^{t}\sin ^{4}{%
\frac{\theta _{b}}{2}}  \nonumber \\
&&-\frac{1}{\sqrt{2}}[\overline{\Gamma }_{\mathit{R}}(0,1)\cos \widetilde{%
\varphi _{b}}+\overline{\Gamma }_{\mathit{I}}(0,1)\sin
\widetilde{\varphi _{b}}]\cos \theta _{2}^{t}\sin {\theta _{b}}%
\sin ^{2}{\frac{\theta _{b}}{2}}
\end{eqnarray}
\begin{equation}
{Im}({\mathbf{\overline{r}_{+-}^{\overline{b}_{R}}} })=-\frac{1}{\sqrt{2}}[%
\overline{\Gamma }_{\mathit{R}}(0,1)\sin \widetilde{\varphi _{b}}-\overline{\Gamma }_{\mathit{I}}(0,1)\cos \widetilde{\varphi _{b}}%
]\sin {\theta _{b}}\sin ^{2}{\frac{\theta _{b}}{2}}
\end{equation}
and $\mathbf{\overline{r}_{+-}^{\overline{b}_{R}}}=(\mathbf{\overline{r}_{-+}^{\overline{b}%
_{R}}})^{\ast }$ .

For the subdominant ${\mathbf{\bar{b}}}_{L}$ decay channel,
\begin{equation}
\mathbf{\bar{R}^{b_{L}}=}\left(
\begin{array}{cc}
\mathbf{\bar{R}^{\bar{b}_{L}}}_{++} & e^{\imath \phi _{2}}\mathbf{\bar{r}^{\bar{b}_{L}}}_{+-} \\
e^{-\imath \phi _{2}}\mathbf{\bar{r}^{\bar{b}_{L}}}_{-+} &
\mathbf{\bar{R}^{\bar{b}_{L}}}_{--}
\end{array}
\right)
\end{equation}
\begin{eqnarray}
\overline{\mathbf{R}}_{++}^{\overline{b}_{L}}
&=&\frac{1}{2}\overline{\Gamma
}^{L}(0,0)\cos ^{2}\frac{\theta _{2}^{t}}{2}\sin ^{2}{\theta _{b}}+%
\overline{\Gamma }^{L}(-1,-1)\sin ^{2}\frac{\theta _{2}^{t}}{2}\cos ^{4}%
{\frac{\theta _{b}}{2}}  \nonumber \\
&&+\frac{1}{\sqrt{2}}[\overline{\Gamma
}_{\mathit{R}}^{L}(0,-1)\cos \widetilde{\varphi
_{b}}-\overline{\Gamma }_{\mathit{I}}^{L}(0,-1)\sin
\widetilde{\varphi _{b}}]\sin \theta _{2}^{t}\sin {\theta _{b}}%
\cos ^{2}{\frac{\theta _{b}}{2}}
\end{eqnarray}
\begin{eqnarray}
\overline{\mathbf{R}}_{--}^{\overline{b}_{L}}
&=&\frac{1}{2}\overline{\Gamma
}^{L}(0,0)\sin ^{2}\frac{\theta _{2}^{t}}{2}\sin ^{2}{\theta _{b}}+%
\overline{\Gamma }^{L}(-1,-1)\cos ^{2}\frac{\theta _{2}^{t}}{2}\cos ^{4}%
{\frac{\theta _{b}}{2}}  \nonumber \\
&&-\frac{1}{\sqrt{2}}[\overline{\Gamma
}_{\mathit{R}}^{L}(0,-1)\cos \widetilde{\varphi
_{b}}-\overline{\Gamma }_{\mathit{I}}^{L}(0,-1)\sin
\widetilde{\varphi _{b}}]\sin \theta _{2}^{t}\sin {\theta _{b}}%
\cos ^{2}{\frac{\theta _{b}}{2}}
\end{eqnarray}
\begin{eqnarray}
\mathit{{Re}}(\mathbf{\overline{r}_{+-}^{\overline{b}_{L}}}) &=&\frac{1}{4}%
\overline{\Gamma }^{L}(0,0)\sin \theta _{2}^{t}\sin ^{2}{\theta
_{b}}-\frac{1}{2}\overline{\Gamma }^{L}(-1,-1)\sin \theta _{2}^{t}\cos ^{4}%
{\frac{\theta _{b}}{2}}  \nonumber \\
&&-\frac{1}{\sqrt{2}}[\overline{\Gamma
}_{\mathit{R}}^{L}(0,-1)\cos \widetilde{\varphi
_{b}}-\overline{\Gamma }_{\mathit{I}}^{L}(0,-1)\sin
\widetilde{\varphi _{b}}]\cos \theta _{2}^{t}\sin {\theta _{b}}%
\cos ^{2}{\frac{\theta _{b}}{2}}
\end{eqnarray}
\begin{equation}
{Im}({\mathbf{\overline{r}_{+-}^{\overline{b}_{L}}} })=-\frac{1}{\sqrt{2}}[%
\overline{\Gamma }_{\mathit{R}}^{L}(0,-1)\sin \widetilde{\varphi _{b}}+%
\overline{\Gamma }_{\mathit{I}}^{L}(0,-1)\cos \widetilde{\varphi
_{b}}]\sin {\theta _{b}}\cos ^{2}{\frac{\theta _{b}}{2}}
\end{equation}
and $\mathbf{\overline{r}_{+-}^{\overline{b}_{L}}}=(\mathbf{\overline{r}_{-+}^{\overline{b}%
_{L}}})^{\ast }$ .
\begin{eqnarray}
\overline{\Gamma }^{L}(0,0) & \equiv &\left| B(0,-1/2)\right| ^{2}, \, \, \overline{%
\Gamma }^{L}(-1,-1)\equiv \left| B(-1,-1/2)\right| ^{2}   \\
\overline{\Gamma }_{\mathit{R}}^{L}(0,-1) &=&\overline{\Gamma }_{\mathit{R}%
}^{L}(-1,0)\equiv \mathit{{Re}}[B(0,-1/2)B(-1,-1/2)^{\ast }]   \\
& \equiv &|B(0,-1/2)||B(-1,-1/2)|\cos \overline{\beta }_{L}   \\
\overline{\Gamma }_{\mathit{I}}^{L}(0,-1) &=&-\overline{\Gamma }_{\mathit{I}%
}^{L}(-1,0)\equiv {Im}[B(0,-1/2)B(-1,-1/2)^{\ast }]   \\
& \equiv &-|B(0,-1/2)||B(-1,-1/2)|\sin \overline{\beta }_{L}
\end{eqnarray}

Sometimes in the derivation, we will denote ${\mathbf{r}_{+-}}
=F_{a}+\imath H_{a}$ and analogously
\newline $
{\mathbf{\overline{r}_{+-}}}=-F_{b}-\imath H_{b}$ \ .  As above,
${b_{L}}$ and $ {b_{R}}$ superscripts on ${\mathbf{r}_{+-}}$, and
on $ F_{a}$ and $ H_{a} $ denote the $\lambda_b = -1/2, 1/2$
contributions, and analogously for ${\mathbf{\overline{r}_{+-}}}$,
$ F_{b}$ and $ H_{b} $.

\subsection{ Start of derivation of BR-S2SC functions}

The general beam-referenced angular distribution in the $(t\bar
t)_{cm}$ is
\begin{equation}
\begin{array}{c}
I(\Theta _B,\Phi _B;\theta _1^t,\phi _1; {\theta
_a},\widetilde{\phi _a};\theta _2^t,\phi
_2;{%
\theta _b},\widetilde{\phi _b}) = \sum_{\lambda _1\lambda
_2\lambda _1^{^{\prime }}\lambda _2^{^{\prime }}}\rho _{\lambda
_1\lambda _2;\lambda _1^{^{\prime }}\lambda _2^{^{\prime
}}}^{\mathtt{prod}}(\Theta _B,\Phi _B) \\
\times R_{\lambda _1\lambda _1^{^{\prime }}} (t\rightarrow
W^{+}b\rightarrow \ldots )
 \overline{ { R } }_{\lambda _2\lambda
_2^{^{\prime }}}(\bar t\rightarrow W^{-}\bar b\rightarrow \ldots )
\end{array}
\end{equation}
where the summations are over the $t_1$ and $\bar{t}_2$
helicities.  The composite decay-density-matrices $R_{\lambda
_1\lambda _1^{^{\prime }}}$ for $t\rightarrow W^{+}b\rightarrow
\ldots $ and ${\overline{R}_{\lambda _2\lambda _2^{^{\prime }}}}$
for $\bar t\rightarrow W^{-}\bar b\rightarrow \ldots $ are given
in the preceding subsection. This formula holds for any of the
above $t \bar{t}$ production channels and for either the
lepton-plus-jets, the dilepton-plus-jets, or the all-jets $t
\bar{t}$ decay channels. The derivation begins in the ``home" or
starting coordinate system $(x_h,y_h,z_h)$ in the
$(t\bar{t})_{c.m.}$ frame.  As shown in Fig. 6-7, the angles
$\Theta _B,\Phi _B$ specify the direction of the incident beam,
the $e$ momentum, or in the case of $p \bar{p} \rightarrow t\bar
tX$, the $q$ momentum arising from the incident $p$ in the $ p\bar
p$. The $t_1$ momentum is chosen to lie along the positive $z_h$
axis. The positive $x_h$ direction is an arbitrary, fixed
perpendicular direction. Because the incident beam is assumed to
be unpolarized, there is no dependence on the associated $\phi_1$
angle after the observable azimuthal angles are specified (see
below). With respect to the normalization of the various BR-S2SC
functions, the $\phi_1$ integration is not explicitly performed in
this paper. With (54) there is an associated differential counting
rate
\begin{equation}
\begin{array}{c}
dN=I(\Theta _B,\Phi _B;\ldots )d(\cos \Theta _B)d\Phi _B d(\cos
\theta _1^t)d\phi _1 \\
d(\cos {\theta _a})d\widetilde{\phi _a}d(\cos \theta _2^t)d\phi _2
d(\cos {\theta _b})d\widetilde{\phi _b}
\end{array}
\end{equation}
where, for full phase space, the cosine of each polar angle ranges
from -1 to 1, and each azimuthal angle ranges over $ 2 \pi $.

For $t\bar{t}$ production by $q \overline{q}$, or $
e\bar{e}\rightarrow t\overline{t}$ by initial unpolarized
particles, the associated production density matrix is derived as
in [5,4].  It is
\begin{eqnarray}
\rho _{\lambda _{1}\lambda _{2};\lambda _{1}^{^{\prime }}\lambda
_{2}^{^{\prime }}}^{\mathtt{prod}} &=&(\frac{1}{s^{2}})e^{\imath
(\lambda ^{\prime }-\lambda )\Phi _{B}}T(\lambda _{1},\lambda
_{2})T^{\ast }(\lambda
_{1}^{^{\prime }},\lambda _{2}^{^{\prime }})  \nonumber \\
&&\times \frac{1}{4}\sum_{s_{1},s_{2}}|\widetilde{T}(s_{1},s_{2})|^{2}d_{%
\lambda s}^{1}(\Theta _{B})d_{\lambda ^{^{\prime }}s}^{1}(\Theta
_{B})
\end{eqnarray}
where $\lambda =\lambda _{1}-\lambda _{2}$, $\lambda ^{^{\prime
}}=\lambda _{1}^{^{\prime }}-\lambda _{2}^{^{\prime }}$, and
$s=s_1 - s_2$. In the body of this paper we concentrate on results
for hadron colliders; formulas for the case of $e \bar{e}$ or $
\mu \bar{\mu} $ production are given in Appendix D. It is
convenient to separate the contributions into three parts,
depending on the roles of the ``helicity-conserving" and
``helicity-flip" $ T(\lambda_1, \lambda_2 )$ amplitudes for $g
\rightarrow t_1 \bar{t}_2$ production.  Relative to the
helicity-conserving amplitudes, the helicity-flip amplitudes are
$(\sqrt{2} m_t/\sqrt{s} )$. We denote by a tilde accent the
corresponding helicity-conserving light-quark $q \bar{q}
\rightarrow g$ annihilation amplitudes. The values $\lambda_{1,2}
= \pm 1/2$ of the arguments of $ T(\lambda_1, \lambda_2 )$ are
denoted by the signs of $\lambda_1$, $\lambda_2$, and likewise for
$ \widetilde{T}(s_1, s_2 )$.

\subsubsection{Helicity-conserving contribution}

The $t_1 \bar{t}_2$ helicity-conserving contribution production
density matrix is
\begin{eqnarray}
\rho _{\lambda _{1}\lambda _{2};\lambda _{1}^{^{\prime }}\lambda
_{2}^{^{\prime }}}^{\mathtt{prod}} &\rightarrow &\delta _{\lambda
_{2},-\lambda _{1}}\delta _{\lambda _{2}^{^{\prime }},-\lambda
_{1}^{^{\prime }}}(\frac{1}{s^{2}})e^{\imath 2(\lambda
_{1}^{^{\prime }}-\lambda _{1})\Phi _{B}}T(\lambda _{1},-\lambda
_{1})T^{\ast }(\lambda
_{1}^{^{\prime }},-\lambda _{1}^{^{\prime }})  \nonumber \\
&&\times \frac{1}{4}\left[ |\widetilde{T}(+-)|^{2}d_{\lambda
1}^{1}(\Theta
_{B})d_{\lambda ^{^{\prime }}1}^{1}(\Theta _{B})+|\widetilde{T}%
(-+)|^{2}d_{\lambda ,-1}^{1}(\Theta _{B})d_{\lambda ^{^{\prime
}},-1}^{1}(\Theta _{B})\right]
\end{eqnarray}
where $\lambda =2\lambda _{1}$ and $\lambda ^{^{\prime }}=2\lambda
_{1}^{^{\prime }}$. \ The angular distribution of (57) has four
different terms which can be labelled as $I_{\lambda ,\lambda
^{^{\prime }}}$ due to the Kronecker $\delta$'s. Explicitly, these
are
\begin{eqnarray}
I_{++}=\frac{1}{4s^{2}}|T(+-)|^{2}{\mathbf{R}_{++}} \overline{{\mathbf{R}}}_{--}%
\left[ |\widetilde{T}(+-)|^{2}\cos ^{4}(\Theta _{B}/2)+|%
\widetilde{T}(-+)|^{2}\sin ^{4}(\Theta _{B}/2)\right]    \\
I_{--}=\frac{1}{4s^{2}}|T(-+)|^{2}{\mathbf{R}_{--}} \overline{{\mathbf{R}}}_{++}%
\left[ |\widetilde{T}(+-)|^{2}\sin ^{4}(\Theta _{B}/2)+|%
\widetilde{T}(-+)|^{2}\cos ^{4}(\Theta _{B}/2)\right]    \\
I_{+-}=\frac{1}{4s^{2}}T(+-)T^{\ast }(-+)e^{-\imath (2\Phi _{R}+\phi )}%
{\mathbf{r}_{+-}} \overline{{\mathbf{r}}}_{-+}  \left[ |\widetilde{T}%
(+-)|^{2}+|\widetilde{T}(-+)|^{2}\right] \cos ^{2}(\Theta
_{B}/2)\sin
^{2}(\Theta _{B}/2)   \\
I_{-+}=\frac{1}{4s^{2}}T(-+)T^{\ast }(+-)e^{\imath (2\Phi _{R}+\phi )}%
{\mathbf{r}_{-+}} \overline{{\mathbf{r}}}_{+-}
\left[ |\widetilde{T}%
(+-)|^{2}+|\widetilde{T}(-+)|^{2}\right] \cos ^{2}(\Theta
_{B}/2)\sin ^{2}(\Theta _{B}/2)
\end{eqnarray}
where the starting angles $\phi _{2}$ and $\Phi _{B}$ have been
replaced by the angles $\phi =\phi _{1}+\phi _{2}$ and $\Phi
_{R}=\Phi _{B}-\phi _{1}$, see Figs. 6-7. \

Two rotations are needed to recast the above expressions in terms
of the angles of the final $(t\overline{t})_{c.m.}$ coordinate
system shown in Figs. 1-2:

\textit{Step1}: We rotate by $\theta _{1}$ so that the new z-axis $\overline{z%
}$ is along the $W_{1}^{+}$ momentum, as shown in Figs. 8-9. \

This replaces the $\Theta _{B},\Phi _{B}$ referencing of the beam
direction by the final polar angle $\theta _{q}$ and an associated
azimuthal $\Phi _{W} $ variable. \ Since this is simply a
coordinate rotation,
\begin{equation}
d(\cos \theta _{q})d\Phi _{W}=d(\cos \Theta _{B})d\Phi _{R}
\end{equation}
The Jacobian is 1, and $\cos \theta _{q}$ and $\Phi _{W}$ have the
usual range for spherical coordinates. \ The formulas for making
this change of variables are:
\begin{eqnarray}
\cos \theta _{q} &=&\cos \theta _{1}\cos \Theta _{B}+\sin \theta
_{1}\sin
\Theta _{B}\cos \Phi _{R}   \\
\sin \theta _{q}\cos \Phi _{W} &=&-\sin \theta _{1}\cos \Theta
_{B}+\cos
\theta _{1}\sin \Theta _{B}\cos \Phi _{R}   \\
\sin \theta _{q}\sin \Phi _{W} &=&\sin \Theta _{B}\sin \Phi _{R}
\end{eqnarray}
and
\begin{equation}
\cos \Theta _{B}=\cos \theta _{1}\cos \theta _{q}-\sin \theta
_{1}\sin \theta _{q}\cos \Phi _{W}
\end{equation}

In Fig. 9, the $W_{2}^{-}$ momentum is at angles $\Theta _{2}$ and
$\Phi _{2}$ . \ Since  $\Theta _{2}=\pi -\psi $, $\Theta _{2}$ can
be replaced by the opening angle $\psi $ between the $W_{1}^{+}$
and\ $W_{2}^{-}$ momenta. \ The opening angle $\psi $ is simply
related to the important angle $\phi =\phi _{1}+\phi _{2}$ between
the $t_{1} $ and $\overline{t}_{2}$ decay planes:  \
\begin{eqnarray}
\cos \psi  &=&-\cos \Theta _{2}=-\cos \theta _{1}\cos \theta
_{2}+\sin
\theta _{1}\sin \theta _{2}\cos \phi  \\
\sin \psi  &=&\sin \Theta _{2}=(1-\cos ^{2}\Theta _{2})^{1/2}
\end{eqnarray}
On the other hand, $ \cos \Phi _{2}$ and $\sin \Phi _{2}$ are
auxiliary variables that appear in the formulas in Appendix C for
transforming the initial beam-referencing spherical angles $\Theta
_{B},\Phi _{R}$ of Figs. 6-7 to the final ones, $\theta _{q},\phi
_{q}$ of Figs. 1-2.
\begin{eqnarray}
\sin \psi \cos \Phi _{2} &=&\sin \theta _{1}\cos \theta _{2}+\cos
\theta
_{1}\sin \theta _{2}\cos \phi    \\
\sin \psi \sin \Phi _{2} &=&\sin \theta _{2}\sin \phi
\end{eqnarray}

\textit{Step 2}: We rotate by $-\Phi _{2}$ about
$\overline{z}=\widehat{z}$ so that the $W_{2}^{-}$ momenta is in
the positive $\widehat{x}$ plane, as shown in Figs. 1-2.

By this rotation,
\begin{equation}
\phi _{q}=\Phi _{W}+\Phi _{2}
\end{equation}
so the Jacobian is 1, and $\phi _{q}$ has the full $2\pi $ range.

By these two steps, the above four helicity-conserving
contributions are expressed in terms of Figs. 1-2:
\begin{eqnarray}
I_{++}+I_{--} &=& \frac{1}{16\,s^{2}} {S}_{q}\left\{ |T(+-)|^{2}{ \mathbf{R}_{++}%
\overline{\mathbf{R}}_{--}} +|T(-+)|^{2}\mathbf{R}_{--}\overline{\mathbf{R}}%
_{++}\right\} \left( 1+\cos ^{2}\Theta _{B}\right)   \nonumber \\
&&+\frac{1}{8 \,s^{2}}{T}_{q}\left\{ |T(+-)|^{2}{ \mathbf{R}_{++}\overline{\mathbf{%
R}}_{--} }
-|T(-+)|^{2}\mathbf{R}_{--}\overline{\mathbf{R}}_{++}\right\} \cos
\Theta _{B}   \\
I_{+-}+I_{-+} &=&-\frac{1}{8 \,s^{2}}{S}_{q}\left\{
\overline{\kappa }\left[ F_{a}F_{b}+H_{a}H_{b}\right]
+\overline{\kappa }^{^{\prime }}\left[
F_{a}H_{b}-H_{a}F_{b}\right] \right\} \sin ^{2}\Theta _{B}\cos
(2\Phi
_{R}+\phi )  \nonumber \\
&&-\frac{1}{8 \,s^{2}}{S}_{q}\left\{ \overline{\kappa }^{^{\prime
}}\left[
F_{a}F_{b}+H_{a}H_{b}\right] -\overline{\kappa }\left[ F_{a}H_{b}-H_{a}F_{b}%
\right] \right\} \sin ^{2}\Theta _{B}\sin (2\Phi _{R}+\phi )
\end{eqnarray}
where
\begin{eqnarray}
S_{q} &=&|\widetilde{T}(+-)|^{2}+|\widetilde{T}(-+)|^{2} \\
T_{q} &=&|\widetilde{T}(+-)|^{2}-|\widetilde{T}(-+)|^{2} \\
\overline{\kappa }+\imath \overline{\kappa }^{^{\prime }}
&=&T(+-)T^{\ast }(-+)
\end{eqnarray}

\subsubsection{Mixed helicity-properties contribution}

The mixed helicity-properties contribution of the
$t_{1}\bar{t}_{2}$ production density matrix is in two parts: The
first part is
\begin{eqnarray}
\rho _{\lambda _{1}\lambda _{2};\lambda _{1}^{^{\prime }}\lambda
_{2}^{^{\prime }}}^{\mathtt{prod}} &\rightarrow &\delta _{\lambda
_{2},\lambda _{1}}\delta _{\lambda _{2}^{^{\prime }},-\lambda
_{1}^{^{\prime }}}(\frac{1}{s^{2}})e^{\imath 2\lambda
_{1}^{^{\prime }}\Phi _{B}}T(\lambda _{1},\lambda _{1})T^{\ast
}(\lambda _{1}^{^{\prime }},-\lambda
_{1}^{^{\prime }})  \nonumber \\
&&\times \frac{1}{4}\left[
|\widetilde{T}(+-)|^{2}d_{0,1}^{1}(\Theta
_{B})d_{\lambda ^{^{\prime }},1}^{1}(\Theta _{B})+|\widetilde{T}%
(-+)|^{2}d_{0,-1}^{1}(\Theta _{B})d_{\lambda ^{^{\prime
}},-1}^{1}(\Theta _{B})\right]
\end{eqnarray}
where $\lambda ^{^{\prime }}=2\lambda _{1}^{^{\prime }}$.

As in the above subsection for the helicity-conserving
contribution, this mixed-helicity properties contribution can be
expressed as the sum of
\begin{eqnarray}
I_{++}^{mA} &=&-\frac{1}{8\sqrt{2}\,s^{2}}(\overline{\eta
}^{+}+\imath
\overline{\eta }^{^{\prime }+}){{\mathbf{R}}_{++}(F_{b}+\imath H_{b})(S}%
_{q}\cos \Theta _{B}+T_{q})\sin \Theta _{B} e^{\imath ( \Phi _{R}
+ \phi )}
\end{eqnarray}
\begin{eqnarray}
I_{--}^{mA} &=&\frac{1}{8\sqrt{2}\,s^{2}}(\overline{\omega
}^{-}+\imath
\overline{\omega }^{^{\prime }-}){{\mathbf{R}}_{--}(F_{b}-\imath H_{b})(S}%
_{q}\cos \Theta _{B}-T_{q})\sin \Theta _{B} e^{- \imath ( \Phi
_{R} + \phi )}
\end{eqnarray}
\begin{eqnarray}
I_{+-}^{mA} &=&-\frac{1}{8\sqrt{2}\,s^{2}}(\overline{\omega
}^{+}+\imath
\overline{\omega }^{^{\prime }+}){(F_{a}+\imath H_{a})}\overline{{{\mathbf{R}%
}}}{_{++}(S}_{q}\cos \Theta _{B}-T_{q})\sin \Theta _{B} e^{-
\imath \Phi _{R} }
\end{eqnarray}
\begin{eqnarray}
I_{-+}^{mA} &=&\frac{1}{8\sqrt{2}\,s^{2}}(\overline{\eta
}^{-}+\imath
\overline{\eta }^{^{\prime }-}){(F_{a}-\imath H_{a})}\overline{{{\mathbf{R}}}%
}{_{--}(S}_{q}\cos \Theta _{B}+T_{q})\sin \Theta _{B} e^{\imath
\Phi _{R} }
\end{eqnarray}
where
\begin{eqnarray}
\overline{\omega }^{+}+\imath \overline{\omega }^{^{\prime }+}
&=&T(++)T^{\ast }(-+) \\
\overline{\omega }^{-}+\imath \overline{\omega }^{^{\prime }-}
&=&T(--)T^{\ast }(-+) \\
\overline{\eta }^{+}+\imath \overline{\eta }^{^{\prime }+}
&=&T(++)T^{\ast
}(+-) \\
\overline{\eta }^{-}+\imath \overline{\eta }^{^{\prime }-}
&=&T(--)T^{\ast }(+-)
\end{eqnarray}

The second part of the $t_{1}\bar{t}_{2}$ mixed
helicity-properties part of the production density matrix is
\begin{eqnarray}
\rho _{\lambda _{1}\lambda _{2};\lambda _{1}^{^{\prime }}\lambda
_{2}^{^{\prime }}}^{\mathtt{prod}} &\rightarrow &\delta _{\lambda
_{2},-\lambda _{1}}\delta _{\lambda _{2}^{^{\prime }},\lambda
_{1}^{^{\prime }}}(\frac{1}{s^{2}})e^{-\imath 2\lambda _{1}\Phi
_{B}}T(\lambda _{1},-\lambda _{1})T^{\ast }(\lambda _{1}^{^{\prime
}},\lambda
_{1}^{^{\prime }})  \nonumber \\
&&\times \frac{1}{4}\left[ |\widetilde{T}(+-)|^{2}d_{\lambda,
1}^{1}(\Theta _{B})d_{0,1}^{1}(\Theta
_{B})+|\widetilde{T}(-+)|^{2}d_{\lambda ,-1}^{1}(\Theta
_{B})d_{0,-1}^{1}(\Theta _{B})\right]
\end{eqnarray}
where $\lambda =2\lambda _{1}$ . \ This mixed-helicity-properties
contribution can be expressed as the sum of
\begin{eqnarray}
I_{++}^{mB} &=&-\frac{1}{8\sqrt{2}\,s^{2}}(\overline{\eta
}^{+}-\imath
\overline{\eta }^{^{\prime }+}){{\mathbf{R}}_{++}(F_{b}-\imath H_{b})(S}%
_{q}\cos \Theta _{B}+T_{q})\sin \Theta _{B} e^{- \imath ( \Phi
_{R} + \phi )}
\end{eqnarray}
\begin{eqnarray}
I_{--}^{mB} &=&\frac{1}{8\sqrt{2}\,s^{2}}(\overline{\omega
}^{-}-\imath
\overline{\omega }^{^{\prime }-}){{\mathbf{R}}_{--}(F_{b}+\imath H_{b})(S}%
_{q}\cos \Theta _{B}-T_{q})\sin \Theta _{B} e^{ \imath ( \Phi _{R}
+ \phi )}
\end{eqnarray}
\begin{eqnarray}
I_{+-}^{mB} &=&\frac{1}{8\sqrt{2}\,s^{2}}(\overline{\eta
}^{-}-\imath
\overline{\eta }^{^{\prime }-}){(F_{a}+\imath H_{a})}\overline{{{\mathbf{R}}}%
}{_{--}[(S}_{q}\cos \Theta _{B}+T_{q})\sin \Theta _{B} e^{- \imath
\Phi _{R} }
\end{eqnarray}
\begin{eqnarray}
I_{-+}^{mB} &=&-\frac{1}{8\sqrt{2}\,s^{2}}(\overline{\omega
}^{+}-\imath
\overline{\omega }^{^{\prime }+}){(F_{a}-\imath H_{a})}\overline{{{\mathbf{R}%
}}}{_{++}(S}_{q}\cos \Theta _{B}-T_{q})\sin \Theta _{B} e^{ \imath
\Phi _{R} }
\end{eqnarray}

\subsubsection{Helicity-flip contribution}

The $t_1 \bar{t}_2$ helicity-flip production density matrix is
\begin{eqnarray}
\rho _{\lambda _{1}\lambda _{2};\lambda _{1}^{^{\prime }}\lambda
_{2}^{^{\prime }}}^{\mathtt{prod}} &\rightarrow &\delta _{\lambda
_{2},\lambda _{1}}\delta _{\lambda _{2}^{^{\prime }},\lambda
_{1}^{^{\prime }}}(\frac{1}{s^{2}})T(\lambda _{1},\lambda
_{1})T^{\ast }(\lambda
_{1}^{^{\prime }},\lambda _{1}^{^{\prime }})  \nonumber \\
&&\times \frac{1}{4}\left[
|\widetilde{T}(+-)|^{2}d_{01}^{1}(\Theta _{B})d_{01}^{1}(\Theta
_{B})+|\widetilde{T}(-+)|^{2}d_{0,-1}^{1}(\Theta
_{B})d_{0,-1}^{1}(\Theta _{B})\right]
\end{eqnarray}
\ This contribution can be expressed as the sum of
\begin{equation}
I_{++}^{m2}+I_{--}^{m2}=\frac{1}{8\,s^{2}}{S}_{q}\left\{ |T(++)|^{2}{\mathbf{%
R}_{++}\overline{\mathbf{R}}_{++}}+|T(--)|^{2}\mathbf{R}_{--}\overline{%
\mathbf{R}}_{--}\right\} \sin ^{2}\Theta _{B}
\end{equation}
and
\begin{eqnarray}
I_{+-}^{m2}+I_{-+}^{m2} &=&\frac{1}{4\,s^{2}}{S}_{q}(\left\{ -\overline{%
\zeta }\left[ F_{a}F_{b}-H_{a}H_{b}\right] +\overline{\zeta }^{^{\prime }}%
\left[ F_{a}H_{b}+H_{a}F_{b}\right] \right\} \cos \phi   \nonumber \\
&&+\left\{ \overline{\zeta }^{^{\prime }}\left[
F_{a}F_{b}-H_{a}H_{b}\right] +\overline{\zeta }\left[
F_{a}H_{b}+H_{a}F_{b}\right] \right\} \sin \phi )\sin ^{2}\Theta
_{B}
\end{eqnarray}
where
\begin{equation}
\overline{\zeta }+\imath \overline{\zeta }^{^{\prime
}}=T(++)T^{\ast }(--)
\end{equation}

For $q \overline{q} \rightarrow t\overline{t}$, in the Jacob-Wick
phase convention, the associated helicity amplitudes are $%
\widetilde{T}(+,-)=\widetilde{T}(-,+)=g,$ the helicity-conserving $%
T(+-)=T(-+)=g,$ and the helicity-flip
$T(++)=T(--)=gm_{t}\sqrt{2/s}$.

\section{ Lepton-plus-Jets Channel:  $ \lambda_b = -1/2$, $\lambda_{\bar{b}} = +1/2$
 \newline Dominance }

From the perspective of specific helicity amplitude tests, one can
use the above results to investigate various BR-S2SC functions for
the lepton-plus-jets channel: \ In this paper, we are interested
in tests for the relative sign of, or for measurement of a
possible non-trivial phase between the $\lambda _{b}=-1/2$
helicity amplitudes for $t\rightarrow W^{+}b $. \ We assume
that the $\lambda _{b}=-1/2$ and $\lambda \overline{_{b}}$ $%
=1/2$ contributions dominate.

\subsection{$t_{1}\rightarrow W_{1}^{+}b\rightarrow (l^{+}\nu
)b$}

For the case $t_{1}\rightarrow W_{1}^{+}b\rightarrow (l^{+}\nu )b$, \ with $%
W_{2}^{-}$ decaying into hadronic jets, we separate the intensity
contributions into two parts: \  ``signal terms''
$\widetilde{I}|_{sig}$ which depend on $\Gamma
_{R}(0,-1)$ and $\Gamma _{I}(0,-1)$, and ``background terms'' $\widetilde{I|}%
_{0}$ which depend on $\Gamma (0,0)$ and $\Gamma (-1,-1)$.\ We use a tilde accent
on $\widetilde{I|}%
_{0}, \ldots $  to denote the integration over the $\theta _{b}$, $%
\widetilde{\phi }_{b}$ variables. This integration gives
\begin{eqnarray}
\int_{-1}^{1}d(\cos \theta _{b})\int_{0}^{2\pi }d\widetilde{\phi }_{b}%
\, \overline{\mathbf{R}}_{++}^{\overline{b}_{R}} &=&\frac{4\pi }{3}[\overline{%
\Gamma }(0,0)\sin ^{2}\frac{\theta _{2}^{t}}{2}+\overline{\Gamma
}(1,1)\cos
^{2}\frac{\theta _{2}^{t}}{2}] \\
\int_{-1}^{1}d(\cos \theta _{b})\int_{0}^{2\pi }d\widetilde{\phi }_{b}%
\, \overline{\mathbf{R}}_{--}^{\overline{b}_{R}} &=&\frac{4\pi }{3}[\overline{%
\Gamma }(0,0)\cos ^{2}\frac{\theta _{2}^{t}}{2}+\overline{\Gamma
}(1,1)\sin ^{2}\frac{\theta _{2}^{t}}{2}]
\end{eqnarray}
\begin{equation}
\int_{-1}^{1}d(\cos \theta _{b})\int_{0}^{2\pi }d\widetilde{\phi }_{b} \, F_{b}^{%
\overline{b}_{R}}=\frac{2\pi }{3}\sin \theta _{2}^{t}[\overline{\Gamma }%
(0,0)-\overline{\Gamma }(1,1)]
\end{equation}
The integration over $H_{b}^{\overline{b}_{R}}$ vanishes.

We find for the helicity-conserving contribution,
\begin{eqnarray}
(\widetilde{I}_{++}+\widetilde{I}_{--})|_{0} &=&\frac{\pi g^{4}}{12s^{2}}%
(1+\cos ^{2}\Theta _{B})  \\
&&\left\{
\begin{array}{c}
\frac{1}{2}\Gamma (0,0)\sin ^{2}\theta _{a}[\overline{\Gamma
}(0,0)(1+\cos \theta _{1}^{t}\cos \theta
_{2}^{t})+\overline{\Gamma }(1,1)(1-\cos \theta
_{1}^{t}\cos \theta _{2}^{t})] \\
+\Gamma (-1,-1)\sin ^{4}\frac{\theta _{a}}{2}[\overline{\Gamma
}(0,0)(1-\cos \theta _{1}^{t}\cos \theta
_{2}^{t})+\overline{\Gamma }(1,1)(1+\cos \theta _{1}^{t}\cos
\theta _{2}^{t})] \nonumber
\end{array}
\right\}
\end{eqnarray}
\begin{eqnarray}
(\widetilde{I}_{++}+\widetilde{I}_{--})|_{sig} &=&\frac{\pi g^{4}}{6\sqrt{2}%
s^{2}}(1+\cos ^{2}\Theta _{B})\sin \theta _{1}^{t}\cos \theta
_{2}^{t}\sin
\theta _{a}\sin ^{2}\frac{\theta _{a}}{2} \\
&&\left\{ -\Gamma _{R}(0,-1)\cos \widetilde{\phi }_{a}+\Gamma
_{I}(0,-1)\sin
\widetilde{\phi }_{a}\right\} [\overline{\Gamma }(0,0)-\overline{\Gamma }%
(1,1)]\nonumber
\end{eqnarray}
\begin{eqnarray}
(\widetilde{I}_{+-}+\widetilde{I}_{-+})|_{0} &=&-\frac{\pi g^{4}}{12s^{2}}%
\sin ^{2}\Theta _{B}\cos (2\Phi _{R}+\phi )\sin \theta
_{1}^{t}\sin \theta
_{2}^{t} \\
&&\left\{ \frac{1}{2}\Gamma (0,0)\sin ^{2}\theta _{a}-\Gamma (-1,-1)\sin ^{4}%
\frac{\theta _{a}}{2}\right\} [\overline{\Gamma }(0,0)-\overline{\Gamma }%
(1,1)]\nonumber
\end{eqnarray}
\begin{eqnarray}
(\widetilde{I}_{+-}+\widetilde{I}_{-+})|_{sig} &=&-\frac{\pi g^{4}}{6\sqrt{2}%
s^{2}}\sin ^{2}\Theta _{B}\sin \theta _{2}^{t}\sin \theta _{a}\sin ^{2}\frac{%
\theta _{a}}{2} [\overline{\Gamma }(0,0)-\overline{\Gamma
}(1,1)] \\
&&\left\{
\begin{array}{c}
\cos (2\Phi _{R}+\phi )\cos \theta _{1}^{t}\left\{ \Gamma
_{R}(0,-1)\cos \widetilde{\phi }_{a}-\Gamma _{I}(0,-1)\sin
\widetilde{\phi }_{a}\right\}
\\
+\sin (2\Phi _{R}+\phi )\left\{ \Gamma _{R}(0,-1)\sin \widetilde{\phi }%
_{a}+\Gamma _{I}(0,-1)\cos \widetilde{\phi }_{a}\right\}
\end{array}
\right\} \nonumber
\end{eqnarray}

For the mixed-helicity contribution, the terms with primed
coefficients [see (82-85)] all vanish. We collect the other
mixed-helicity contributions in real sums:
\begin{eqnarray}
\widetilde{I}^{m(\overline{\omega }^{+}+\overline{\eta }^{-})}|_{0} &=&\frac{%
\pi g^{4}m_{t}}{3s^{2}\sqrt{s}}\sin \Theta _{B}\cos \Theta
_{B}\cos \Phi
_{R}\sin \theta _{1}^{t}\cos \theta _{2}^{t} \\
&&\left\{ \frac{1}{2}\Gamma (0,0)\sin ^{2}\theta _{a}-\Gamma (-1,-1)\sin ^{4}%
\frac{\theta _{a}}{2}\right\} [\overline{\Gamma }(0,0)-\overline{\Gamma }%
(1,1)]\nonumber
\end{eqnarray}
\begin{eqnarray}
\widetilde{I}^{m(\overline{\omega }^{+}+\overline{\eta }^{-})}|_{sig} &=&%
\frac{\sqrt{2}\pi g^{4}m_{t}}{3s^{2}\sqrt{s}}\sin \Theta _{B}\cos
\Theta
_{B}\cos \theta _{2}^{t}\sin \theta _{a}\sin ^{2}\frac{\theta _{a}}{2} \\
&&\left\{
\begin{array}{c}
\cos \theta _{1}^{t}\left\{ \Gamma _{R}(0,-1)\cos \widetilde{\phi }%
_{a}-\Gamma _{I}(0,-1)\sin \widetilde{\phi }_{a}\right\} \cos \Phi _{R} \\
+\left\{ \Gamma _{R}(0,-1)\sin \widetilde{\phi }_{a}+\Gamma
_{I}(0,-1)\cos \widetilde{\phi }_{a}\right\} \sin \Phi _{R}
\end{array}
\right\} [\overline{\Gamma }(0,0)-\overline{\Gamma
}(1,1)]\nonumber
\end{eqnarray}
\begin{eqnarray}
\widetilde{I}^{m(\overline{\omega }^{-}+\overline{\eta }^{+})}|_{0} &=&-%
\frac{\pi g^{4}m_{t}}{3s^{2}\sqrt{s}}\sin \Theta _{B}\cos \Theta
_{B}\cos
(\Phi _{R}+\phi )\cos \theta _{1}^{t}\sin \theta _{2}^{t} \\
&&\left\{ \frac{1}{2}\Gamma (0,0)\sin ^{2}\theta _{a}-\Gamma (-1,-1)\sin ^{4}%
\frac{\theta _{a}}{2}\right\} [\overline{\Gamma }(0,0)-\overline{\Gamma }%
(1,1)]\nonumber
\end{eqnarray}
\begin{eqnarray}
\widetilde{I}^{m(\overline{\omega }^{-}+\overline{\eta }^{+})}|_{sig} &=&%
\frac{\sqrt{2}\pi g^{4}m_{t}}{3s^{2}\sqrt{s}}\sin \Theta _{B}\cos
\Theta _{B}\cos (\Phi _{R}+\phi )\sin \theta _{1}^{t}\sin \theta
_{2}^{t}\sin
\theta _{a}\sin ^{2}\frac{\theta _{a}}{2} \\
&&\left\{ \Gamma _{R}(0,-1)\cos \widetilde{\phi }_{a}-\Gamma
_{I}(0,-1)\sin
\widetilde{\phi }_{a}\right\} [\overline{\Gamma }(0,0)-\overline{\Gamma }%
(1,1)]\nonumber
\end{eqnarray}

The helicity-flip contributions are
\begin{eqnarray}
(\widetilde{I}_{++}^{m2}+\widetilde{I}_{--}^{m2})|_{0}
&=&\frac{\pi
g^{4}m_{t}^{2}}{3s^{3}}\sin ^{2}\Theta _{B} \\
&&\left\{
\begin{array}{c}
\frac{1}{2}\Gamma (0,0)\sin ^{2}\theta _{a}[\overline{\Gamma
}(0,0)(1-\cos \theta _{1}^{t}\cos \theta
_{2}^{t})+\overline{\Gamma }(1,1)(1+\cos \theta
_{1}^{t}\cos \theta _{2}^{t})] \\
+\Gamma (-1,-1)\sin ^{4}\frac{\theta _{a}}{2}[\overline{\Gamma
}(0,0)(1+\cos \theta _{1}^{t}\cos \theta
_{2}^{t})+\overline{\Gamma }(1,1)(1-\cos \theta _{1}^{t}\cos
\theta _{2}^{t})]
\end{array}
\right\} \nonumber
\end{eqnarray}
\begin{eqnarray}
(\widetilde{I}_{++}^{m2}+\widetilde{I}_{--}^{m2})|_{sig} &=&\frac{\sqrt{2}%
\pi g^{4}m_{t}^{2}}{3s^{3}}\sin ^{2}\Theta _{B}\sin \theta
_{1}^{t}\cos
\theta _{2}^{t}\sin \theta _{a}\sin ^{2}\frac{\theta _{a}}{2} \\
&&\left\{ \Gamma _{R}(0,-1)\cos \widetilde{\phi }_{a}-\Gamma
_{I}(0,-1)\sin
\widetilde{\phi }_{a}\right\} [\overline{\Gamma }(0,0)-\overline{\Gamma }%
(1,1)]\nonumber
\end{eqnarray}
\begin{eqnarray}
(\widetilde{I}_{+-}^{m2}+\widetilde{I}_{-+}^{m2})|_{0}
&=&\frac{\pi g^{4}m_{t}^{2}}{3s^{3}}\sin ^{2}\Theta _{B}\cos \phi
\sin \theta
_{1}^{t}\sin \theta _{2}^{t} \\
&&\left\{ -\frac{1}{2}\Gamma (0,0)\sin ^{2}\theta _{a}+\Gamma
(-1,-1)\sin
^{4}\frac{\theta _{a}}{2}\right\} [\overline{\Gamma }(0,0)-\overline{\Gamma }%
(1,1)]\nonumber
\end{eqnarray}
\begin{eqnarray}
(\widetilde{I}_{+-}^{m2}+\widetilde{I}_{-+}^{m2})|_{sig} &=&\frac{\sqrt{2}%
\pi g^{4}m_{t}^{2}}{3s^{3}}\sin ^{2}\Theta _{B}\sin \theta
_{2}^{t}\sin
\theta _{a}\sin ^{2}\frac{\theta _{a}}{2} \\
&&\left\{
\begin{array}{c}
\cos \phi \cos \theta _{1}^{t}\left\{ -\Gamma _{R}(0,-1)\cos
\widetilde{\phi
}_{a}+\Gamma _{I}(0,-1)\sin \widetilde{\phi }_{a}\right\}  \\
+\sin \phi \left\{ \Gamma _{R}(0,-1)\sin \widetilde{\phi
}_{a}+\Gamma _{I}(0,-1)\cos \widetilde{\phi }_{a}\right\}
\end{array}
\right\} [\overline{\Gamma }(0,0)-\overline{\Gamma
}(1,1)]\nonumber
\end{eqnarray}

\subsection{$\overline{t}_{2}\rightarrow W_{2}^{-}%
\overline{b}\rightarrow (l^{-}\bar{\nu} )\overline{b}$}

For the\ $CP$-conjugate process $\overline{t}_{2}\rightarrow W_{2}^{-}%
\overline{b}\rightarrow (l^{-}\bar{\nu} )\overline{b}$, \ with
$W_{1}^{+}$ decaying into hadronic jets, we similarly separate the
contributions: \
``signal terms'' $\widetilde{\overline{I}|}_{sig}$depending on $\overline{%
\Gamma }_{R}(0,1)$ and $\overline{\Gamma }_{I}(0,1)$, and
``background
terms'' $\widetilde{\overline{I}}|_{0}$ depending on $\overline{\Gamma }%
(0,0) $ and $\overline{\Gamma }(1,1)$. \ The integration over
$\theta _{a}$, $\widetilde{\phi }_{a}$ gives
\begin{equation}
\int_{-1}^{1}d(\cos \theta _{a})\int_{0}^{2\pi }d\widetilde{\phi }_{a}{%
\, \mathbf{R}_{++}^{b_{L}}}=\frac{4\pi }{3}[\Gamma (0,0)\cos
^{2}\frac{\theta _{1}^{t}}{2}+\Gamma (-1,-1)\sin ^{2}\frac{\theta
_{1}^{t}}{2}]
\end{equation}
\begin{equation}
\int_{-1}^{1}d(\cos \theta _{a})\int_{0}^{2\pi }d\widetilde{\phi }_{a}{%
\, \mathbf{R}_{--}^{b_{L}}}=\frac{4\pi }{3}[\Gamma (0,0)\sin
^{2}\frac{\theta _{1}^{t}}{2}+\Gamma (-1,-1)\cos ^{2}\frac{\theta
_{1}^{t}}{2}]
\end{equation}
\begin{equation}
\int_{-1}^{1}d(\cos \theta _{a})\int_{0}^{2\pi }d\widetilde{\phi }%
_{a} \, F_{a}^{b_{L}}=\frac{2\pi }{3}\sin \theta _{1}^{t}[\Gamma
(0,0)-\Gamma (-1,-1)]
\end{equation}
The integration over $H_{a}^{b_{L}}$ vanishes.

We find for the helicity-conserving contribution,
\begin{eqnarray}
(\widetilde{\overline{I}}_{++}+\widetilde{\overline{I}}_{--})|_{0} &=&\frac{%
\pi g^{4}}{12s^{2}}(1+\cos ^{2}\Theta _{B}) \\
&&\left\{
\begin{array}{c}
\frac{1}{2}\overline{\Gamma }(0,0)\sin ^{2}\theta _{b}[\Gamma
(0,0)(1+\cos \theta _{1}^{t}\cos \theta _{2}^{t})+\Gamma
(-1,-1)(1-\cos \theta
_{1}^{t}\cos \theta _{2}^{t})] \\
+\overline{\Gamma }(1,1)\sin ^{4}\frac{\theta _{b}}{2}[\Gamma
(0,0)(1-\cos \theta _{1}^{t}\cos \theta _{2}^{t})+\Gamma
(-1,-1)(1+\cos \theta _{1}^{t}\cos \theta _{2}^{t})]
\end{array}
\right\} \nonumber
\end{eqnarray}
\begin{eqnarray}
(\widetilde{\overline{I}}_{++}+\widetilde{\overline{I}}_{--})|_{sig} &=&%
- \frac{\pi g^{4}}{6\sqrt{2}s^{2}}(1+\cos ^{2}\Theta _{B})\cos
\theta
_{1}^{t}\sin \theta _{2}^{t}\sin \theta _{b}\sin ^{2}\frac{\theta _{b}}{2} \\
&&\left\{ \overline{\Gamma }_{R}(0,1)\cos \widetilde{\phi }_{b}+ \overline{%
\Gamma }_{I}(0,1)\sin \widetilde{\phi }_{b}\right\} [\Gamma
(0,0)-\Gamma (-1,-1)]\nonumber
\end{eqnarray}
\begin{eqnarray}
(\widetilde{\overline{I}}_{+-}+\widetilde{\overline{I}}_{-+})|_{0} &=&-\frac{%
\pi g^{4}}{12s^{2}}\sin ^{2}\Theta _{B}\cos (2\Phi _{R}+\phi )\sin
\theta
_{1}^{t}\sin \theta _{2}^{t} \\
&&\left\{ \frac{1}{2}\overline{\Gamma }(0,0)\sin ^{2}\theta _{b}-\overline{%
\Gamma }(1,1)\sin ^{4}\frac{\theta _{b}}{2}\right\} [\Gamma
(0,0)-\Gamma (-1,-1)]\nonumber
\end{eqnarray}
\begin{eqnarray}
(\widetilde{\overline{I}}_{+-}+\widetilde{\overline{I}}_{-+})|_{sig} &=&-%
\frac{\pi g^{4}}{6\sqrt{2}s^{2}}\sin ^{2}\Theta _{B}\sin \theta
_{1}^{t}\sin
\theta _{b}\sin ^{2}\frac{\theta _{b}}{2} \, [\Gamma (0,0)-\Gamma (-1,-1)]  \\
&&\left\{
\begin{array}{c}
\cos (2\Phi _{R}+\phi )\cos \theta _{2}^{t}\left\{ \overline{\Gamma }%
_{R}(0,1)\cos \widetilde{\phi }_{b}+\overline{\Gamma
}_{I}(0,1)\sin
\widetilde{\phi }_{b}\right\}  \\
-\sin (2\Phi _{R}+\phi )\left\{ \overline{\Gamma }_{R}(0,1)\sin \widetilde{%
\phi }_{b}-\overline{\Gamma }_{I}(0,1)\cos \widetilde{\phi
}_{b}\right\}
\end{array}
\right\} \nonumber
\end{eqnarray}

The mixed-helicity contributions are
\begin{eqnarray}
\widetilde{\overline{I}}^{m(\overline{\omega }^{+}+\overline{\eta }%
^{-})}|_{0} &=&\frac{\pi g^{4}m_{t}}{3s^{2}\sqrt{s}}\sin \Theta
_{B}\cos
\Theta _{B}\cos \Phi _{R}\sin \theta _{1}^{t}\cos \theta _{2}^{t} \\
&&\left\{ \frac{1}{2}\overline{\Gamma }(0,0)\sin ^{2}\theta _{b}-\overline{%
\Gamma }(1,1)\sin ^{4}\frac{\theta _{b}}{2}\right\} [{\Gamma }(0,0)-%
{\Gamma }(-1,-1)]\nonumber
\end{eqnarray}
\begin{eqnarray}
\widetilde{\overline{I}}^{m(\overline{\omega }^{+}+\overline{\eta }%
^{-})}|_{sig} &=&-\frac{\sqrt{2}\pi
g^{4}m_{t}}{3s^{2}\sqrt{s}}\sin \Theta _{B}\cos \Theta _{B}\cos
\Phi _{R}\sin \theta _{1}^{t}\sin \theta
_{2}^{t}\sin \theta _{b}\sin ^{2}\frac{\theta _{b}}{2} \\
&&\left\{ \overline{\Gamma }_{R}(0,1)\cos \widetilde{\phi }_{b}+\overline{%
\Gamma }_{I}(0,1)\sin \widetilde{\phi }_{b}\right\} [\Gamma
(0,0)-\Gamma (-1,-1)]\nonumber
\end{eqnarray}
\begin{eqnarray}
\widetilde{\overline{I}}^{m(\overline{\omega }^{-}+\overline{\eta }%
^{+})}|_{0} &=&-\frac{\pi g^{4}m_{t}}{3s^{2}\sqrt{s}}\sin \Theta
_{B}\cos
\Theta _{B}\cos (\Phi _{R}+\phi )\cos \theta _{1}^{t}\sin \theta _{2}^{t} \\
&&\left\{ \frac{1}{2}\overline{\Gamma }(0,0)\sin ^{2}\theta _{b}-\overline{%
\Gamma }(1,1)\sin ^{4}\frac{\theta _{b}}{2}\right\} [\Gamma
(0,0)-\Gamma (-1,-1)]\nonumber
\end{eqnarray}
\begin{eqnarray}
\widetilde{\overline{I}}^{m(\overline{\omega }^{-}+\overline{\eta }%
^{+})}|_{sig} &=& - \frac{\sqrt{2}\pi
g^{4}m_{t}}{3s^{2}\sqrt{s}}\sin \Theta
_{B}\cos \Theta _{B}\cos \theta _{1}^{t}\sin \theta _{b}\sin ^{2}\frac{%
\theta _{b}}{2} \\
&&\left\{
\begin{array}{c}
\cos \theta _{2}^{t}\left\{ \overline{\Gamma }_{R}(0,1)\cos \widetilde{\phi }%
_{b}+\overline{\Gamma }_{I}(0,1)\sin \widetilde{\phi }_{b}\right\}
\cos
(\Phi _{R}+\phi ) \\
+\left\{ -\overline{\Gamma }_{R}(0,1)\sin \widetilde{\phi }_{b}+\overline{%
\Gamma }_{I}(0,1)\cos \widetilde{\phi }_{b}\right\} \sin (\Phi
_{R}+\phi )
\end{array}
\right\} [\Gamma (0,0)-\Gamma (-1,-1)]\nonumber
\end{eqnarray}

The helicity-flip contributions are
\begin{eqnarray}
(\widetilde{\overline{I}}_{++}^{m2}+\widetilde{\overline{I}}_{--}^{m2})|_{0}
&=&\frac{\pi g^{4}m_{t}^{2}}{3s^{3}}\sin ^{2}\Theta _{B} \\
&&\left\{
\begin{array}{c}
\frac{1}{2}\overline{\Gamma }(0,0)\sin ^{2}\theta _{b}[\Gamma
(0,0)(1-\cos \theta _{1}^{t}\cos \theta _{2}^{t})+\Gamma
(-1,-1)(1+\cos \theta
_{1}^{t}\cos \theta _{2}^{t})] \\
+\overline{\Gamma }(1,1)\sin ^{4}\frac{\theta _{b}}{2}[\Gamma
(0,0)(1+\cos \theta _{1}^{t}\cos \theta _{2}^{t})+\Gamma
(-1,-1)(1-\cos \theta _{1}^{t}\cos \theta _{2}^{t})]
\end{array}
\right\} \nonumber
\end{eqnarray}
\begin{eqnarray}
(\widetilde{\overline{I}}_{++}^{m2}+\widetilde{\overline{I}}%
_{--}^{m2})|_{sig} &=&\frac{\sqrt{2}\pi
g^{4}m_{t}^{2}}{3s^{3}}\sin ^{2}\Theta _{B}\cos \theta
_{1}^{t}\sin \theta _{2}^{t}\sin \theta _{b}\sin
^{2}\frac{\theta _{b}}{2} \\
&&\left\{ \overline{\Gamma }_{R}(0,1)\cos \widetilde{\phi }_{b}+\overline{%
\Gamma }_{I}(0,1)\sin \widetilde{\phi }_{b}\right\} [\Gamma
(0,0)-\Gamma (-1,-1)]  \nonumber
\end{eqnarray}
\begin{eqnarray}
(\widetilde{\overline{I}}_{+-}^{m2}+\widetilde{\overline{I}}_{-+}^{m2})|_{0}
&=&\frac{\pi g^{4}m_{t}^{2}}{3s^{3}}\sin ^{2}\Theta _{B}\cos \phi
\sin
\theta _{1}^{t}\sin \theta _{2}^{t} \\
&&\left\{ -\frac{1}{2}\overline{\Gamma }(0,0)\sin ^{2}\theta _{b}+\overline{%
\Gamma }(1,1)\sin ^{4}\frac{\theta _{b}}{2}\right\} [\Gamma
(0,0)-\Gamma (-1,-1)]\nonumber
\end{eqnarray}
\begin{eqnarray}
(\widetilde{\overline{I}}_{+-}^{m2}+\widetilde{\overline{I}}%
_{-+}^{m2})|_{sig} &=& - \frac{\sqrt{2}\pi
g^{4}m_{t}^{2}}{3s^{3}}\sin
^{2}\Theta _{B}\sin \theta _{1}^{t}\sin \theta _{b}\sin ^{2}\frac{\theta _{b}%
}{2} \\
&&\left\{
\begin{array}{c}
\cos \phi \cos \theta _{2}^{t}\left\{ \overline{\Gamma
}_{R}(0,1)\cos
\widetilde{\phi }_{b}+\overline{\Gamma }_{I}(0,1)\sin \widetilde{\phi }%
_{b}\right\}  \\
-\sin \phi \left\{ \overline{\Gamma }_{R}(0,1)\sin \widetilde{\phi }_{b}-%
\overline{\Gamma }_{I}(0,1)\cos \widetilde{\phi }_{b}\right\}
\end{array}
\right\} [\Gamma (0,0)-\Gamma (-1,-1)]\nonumber
\end{eqnarray}

\subsection{$\Gamma( \lambda_W, {\lambda_W}^{'} )$ tests versus
angular dependence}

\bigskip In summary, with beam-referencing, for the $t_{1}\rightarrow
W_{1}^{+}b\rightarrow (l^{+}{\nu })b$ case there are six
``background terms'' depending on  $\Gamma (0,0)$ and $\Gamma
(-1,-1)$, and also six ``signal terms'' depending on $\Gamma
_{R,I}(0,-1)$. \ As a consequence of Lorentz invariance, there are
associated kinematic factors with simple angular dependence which
can be used to isolate and measure these four $\Gamma ^{\prime
}$s:

(i) $\theta _{a}$ polar-angle dependence:

The coefficients of  $%
\Gamma (0,0) \Big / \Gamma (-1,-1)\Big / \Gamma _{R,I}(0,-1)$ vary
relatively as the $W$-decay \newline $d_{m m^{\prime}} ^1 (
\theta_a )$-squared-intensity-ratios
\begin{eqnarray}
\frac{1}{2}\sin ^{2}\theta _{a}\Big / \left[ \sin ^{4}\frac{\theta _{a}}{2}%
\right] \Big / \left\{ \frac{1}{\sqrt{2}}\sin \theta _{a}\sin ^{2}\frac{%
\theta _{a}}{2}\right\} = \nonumber \\
2(1+\cos \theta _{a}) \Big / \left[ 1-\cos \theta
_{a}\right] \Big / \left\{ \sqrt{2(1+\cos \theta _{a})(1-\cos \theta _{a})}%
=\sqrt{2}\sin \theta _{a}\right\}
\end{eqnarray}

(ii) $\phi _{a}$ azimuthal-angle dependence in the ``signal
terms'' [ or $\widetilde{\phi }_{a}$ dependence if $\bar{t}_{2}$ is used to specify the $%
0^{o}$ direction] :

The coefficients of $\Gamma _{R}(0,-1) \Big / \Gamma _{I}(0,-1)$ \
vary as
\begin{equation}
\cos \phi _{a} \Big / \sin \phi _{a}
\end{equation}
in each of the signal terms.  However, in three terms there are
also $ \Gamma _{R,I}(0,-1)$'s with the opposite association of
these $\cos \phi _{a}$, $ \sin \phi _{a}$ factors. This opposite
association
occurs in $(\widetilde{I}_{+-}+\widetilde{I}_{-+})|_{sig}$, $\widetilde{I}^{m(%
\overline{\omega }^{+}+\overline{\eta }^{-})}|_{sig}$, and $(\widetilde{I}%
_{+-}^{m2}+\widetilde{I}_{-+}^{m2})|_{sig}$, along with a different $%
\Phi _{R}$ and $\phi $ dependence which might be useful
empirically in separation from the terms with the normal $\phi
_{a} $ association.

To reduce the number of angles, we integrate out the two
beam-referencing angles, and also $\phi $:
\begin{eqnarray}
\widetilde{\mathcal{F}}_{i} \equiv \int_{0}^{2\pi }d\phi
\int_{-1}^{1}d(\cos \Theta _{B})\int_{0}^{2\pi }d\Phi
_{R}\widetilde{I}_{i}
\end{eqnarray}
This yields four-angle S2SC functions.

In terms of $ K $ defined in (17), the four-angle distribution $\{
\theta _{1}^{t}$, $\theta _{2}^{t}$, $ \theta _{a}$, $\phi _{a}
\}$ is
\begin{eqnarray}
\widetilde{ \mathcal{F} }  {|}_{0} &=&\frac{8\pi ^{3}g^{4}}{9s^{2}}(1+%
\frac{2m_{t}^{2}}{s}) \\
&&\left\{
\begin{array}{c}
\frac{1}{2}\overline{\Gamma }(0,0)\sin ^{2}\theta _{a}[\overline{\Gamma }%
(0,0)(1+K\cos \theta _{1}^{t}\cos \theta _{2}^{t})+\overline{\Gamma }%
(1,1)(1-K\cos \theta _{1}^{t}\cos \theta _{2}^{t})] \\
+\overline{\Gamma }(-1,-1)\sin ^{4}\frac{\theta _{a}}{2}[\overline{\Gamma }%
(0,0)(1-K\cos \theta _{1}^{t}\cos \theta _{2}^{t})+\overline{\Gamma }%
(1,1)(1+K\cos \theta _{1}^{t}\cos \theta _{2}^{t})] \nonumber
\end{array}
\right\}
\end{eqnarray}
\begin{eqnarray}
{\widetilde{\mathcal{F}}} {|}_{sig} &=&{-}\frac{8\sqrt{2}\pi
^{3}g^{4}}{9s^{2}}(1+\frac{2m_{t}^{2}}{{s}})\cos \theta
_{2}^{t}K\sin
\theta _{1}^{t}\sin \theta _{a}\sin ^{2}\frac{\theta _{a}}{2} \\
&&\left\{ \Gamma _{R}(0,-1)\cos \phi _{a}-\Gamma _{I}(0,-1)\sin
\phi _{a}\right\} [\overline{\Gamma }(0,0)-\overline{\Gamma
}(1,1)] \nonumber
\end{eqnarray}
The terms in these expressions arise from the helicity-conserving
$(\widetilde{{I}}_{++}+\widetilde{{I}}_{--})$, and from the
helicity-flip
$(\widetilde{{I}}_{++}^{m2}+\widetilde{{I}}_{--}^{m2}) $. In each
case there are contributions to both background and signal parts.

Without the integration over $\phi$, there is a contribution to
both the background and signal parts from the helicity-flip
$(\widetilde{{I}}_{+-}^{m2}+\widetilde{{I}}_{-+}^{m2})$ of
(108-9).   This additional contribution has both the normal and
opposite $\phi_a$ dependence as discussed above in (ii).  It will
be fundamentally significant to empirically demonstrate in both
$\cos{\phi}$ and $\sin{\phi}$ the presence of this contribution to
the spin-correlation because it arises completely from the
combination of $t_1$-quark L-R interference and
$\bar{t}_2$-antiquark L-R interference [see (93)].  Without the
$\phi$ dependence, in the above four-angle function (128-9) there
is no contribution from the off-diagonal elements of the $\lambda
_{b}=-1/2$ and $\lambda \overline{_{b}}$ $%
=1/2$ sequential decay matrices (25) and (39).

For the $CP$-conjugate case in terms of $\{ \theta _{2}^{t}$,
$\theta _{1}^{t}$, $\theta _{b}$, $\phi _{b} \}$, the analogous
four-angle distributions are\
\begin{eqnarray}
\widetilde{ \overline{ \mathcal{F} }}  {|}_{0} &=&\frac{8\pi ^{3}g^{4}}{%
9s^{2}}(1+\frac{2m_{t}^{2}}{{s}}) \\
&&\left\{
\begin{array}{c}
\frac{1}{2}{\overline{\Gamma }}(0,0)\sin ^{2}\theta _{b}[\Gamma
(0,0)(1+K\cos \theta _{1}^{t}\cos \theta _{2}^{t})+\Gamma
(-1,-1)(1-K\cos
\theta _{1}^{t}\cos \theta _{2}^{t})] \\
+{\overline{\Gamma }}(1,1)\sin ^{4}\frac{\theta _{b}}{2}[\Gamma
(0,0)(1-K\cos \theta _{1}^{t}\cos \theta _{2}^{t})+\Gamma
(-1,-1)(1+K\cos \theta _{1}^{t}\cos \theta _{2}^{t})] \nonumber
\end{array}
\right\}
\end{eqnarray}
\begin{eqnarray}
\widetilde{ \overline{ \mathcal{F} }}  {|}_{sig}
 &=& - \frac{8 \sqrt{2}\pi ^{3}g^{4}}{%
9s^{2}} (1+\frac{2m_{t}^{2}}{s})\cos \theta _{1}^{t}K\sin \theta
_{2}^{t}\sin \theta _{b}\sin ^{2}\frac{\theta _{b}}{2}
\\
&&\left\{ \overline{\Gamma }_{R}(0,1)\cos \phi _{b}+\overline{\Gamma }%
_{I}(0,1)\sin \phi _{b}\right\} [\Gamma (0,0)-\Gamma (-1,-1)]
\nonumber
\end{eqnarray}

The still simpler three-angle distributions, which were discussed
in the introduction section, then follow if the $\cos \theta
_{1}^{t}$ integration is also performed ${\mathcal{F}}_{i}{\mathcal{\equiv }}\int_{-1}^{1}d(\cos \theta _{1}^{t})%
\widetilde{\mathcal{F}}_{i}$:
\begin{equation}
{ \mathcal{F|}} _{0}{\mathcal{=}}\frac{16\pi ^{3}g^{4}}{9s^{2}}(1+\frac{2m_{t}^{2}%
}{s})\left\{ \frac{1}{2}\Gamma (0,0)\sin ^{2}\theta _{a}+\Gamma
(-1,-1)\sin ^{4}\frac{\theta _{a}}{2}\right\} [\overline{\Gamma }(0,0)+%
\overline{\Gamma }(1,1)]
\end{equation}
\begin{eqnarray}
{\mathcal{F|}}_{sig} &=&{-}\frac{8\pi ^{4}g^{4}}{9s^{2}}(1-%
\frac{2m_{t}^{2}}{s})\cos \theta _{2}^{t} \frac{1}{\sqrt{2}} \sin \theta _{a}\sin ^{2}\frac{%
\theta _{a}}{2} \\
&&\left\{ \Gamma _{R}(0,-1)\cos \phi _{a}-\Gamma _{I}(0,-1)\sin
\phi _{a}\right\} [\overline{\Gamma }(0,0)-\overline{\Gamma
}(1,1)] \nonumber
\end{eqnarray}

The analogous three-angle S2SC function for the $CP$-conjugate $%
\overline{t}_{2}\rightarrow W_{2}^{-}\overline{b}\rightarrow (l^{-}\nu )%
\overline{b}$ is
\begin{equation}
\overline{ { \mathcal{F|}}} _{0}  {\mathcal{=}} \frac{16\pi ^{3}g^{4}}{9s^{2}%
}(1+\frac{2m_{t}^{2}}{s})\left\{ \frac{1}{2}\overline{\Gamma
}(0,0)\sin
^{2}\theta _{b}+\overline{\Gamma }(1,1)\sin ^{4}\frac{\theta _{b}}{2}%
\right\} [\Gamma (0,0)+\Gamma (-1,-1)]
\end{equation}
\begin{eqnarray}
\overline{ { \mathcal{F|}}} _{sig} {\mathcal{=}} -
\frac{8\pi^{4}g^{4}}{9s^{2}}(1-\frac{2m_{t}^{2}}{s})\cos
\theta _{1}^{t}   \frac{1}{\sqrt{2}}  \sin\theta _{b}\sin ^{2}\frac{\theta _{b}}{2} \\
\left\{ \overline{\Gamma }_{R}(0,1)\cos \phi _{b}+\overline{\Gamma }%
_{I}(0,1)\sin \phi _{b}\right\} [\Gamma (0,0)-\Gamma (-1,-1)]
\nonumber
\end{eqnarray}

\section{Discussion }

In the above derivation of general BR-S2SC functions, in part for
greater generality, we include beam-referencing.  At hadron
colliders, beam-referencing may be useful in some applications. In
the case of $e \bar{e} $-production, it would probably be useful
in investigating possible anomalous initial-state-with-final-state
couplings in the $ t_1 \bar{t}_2$ production process.  However,
the simple three-angle formulas reported in the introduction
section do not make use of beam-referencing.  Given the conceptual
simplicity of the helicity formulation for $q \overline{q} , \,
{\mathrm{ or }} \,  \, e\bar{e}\rightarrow
t\overline{t}\rightarrow (W^{+}b)(W^{-}%
\overline{b}) \rightarrow \cdots $, such non-beam-referenced
functions are ideal for tests of the moduli and phases of the four
$t \rightarrow W^+ b$ helicity amplitudes. While usage of direct
boosts from the $(t \bar{t})_{c.m.}$ frame to the $W^+$ or $W^-$
rest frames might be useful for some purposes, from the
perspective of this BR-S2SC helicity formulation, such boosts will
be an unnecessary complication. The boosts introduce additional
Wigner rotations which obscure the overall simplicity of the
helicity formulation which distinctly separates the different
physics stages of the $ t \bar{t}$ production and decay sequences.

In this paper we separate the $\lambda_b = -1/2$ contributions
from the $\lambda_b = 1/2$ contributions.  To display the
$W$-boson polarization and longitudinal-transverse interference
effects, we introduce a transparent $\Gamma^{\lambda_b} ( \lambda
_W,\lambda _W^{^{\prime }} ) $ notation.  Appendix B relates this
notation to the helicity parameters notation used in [5, 11,
12,14]. At the present time, the $\lambda_b = -1/2$ amplitudes do
indeed appear to dominate in the $ t \rightarrow W^+ b $ decay
mode and so the present paper's $\Gamma^{\lambda_b} ( \lambda
_W,\lambda _W^{^{\prime }} ) $ notation is very appropriate.  At a
later date, in higher precision experiments where effects from all
four of the decay amplitudes must be carefully considered, the
helicity parameters notation might be useful.  It is more
analogous to the notation of the Michel-parameters which continue
to be used in muon decay data analysis. On the other hand, in the
context of a characterization of fundamental ``particle
properties", the present $\Gamma^{\lambda_b} ( \lambda _W,\lambda
_W^{^{\prime }} ) $ notation is a simple way to precisely specify
polarized-partial-width measurements, including $W$-boson
longitudinal-transverse interference. Since the $t \rightarrow W^+
b$ decay channel will first be investigated at hadron colliders,
such measurements will be of channel polarized-partial-width
branching ratios
\begin{eqnarray}
B^{\lambda_b} ( \lambda _W,\lambda _W^{^{\prime }} ) =
\Gamma^{\lambda_b} ( \lambda _W,\lambda _W^{^{\prime }} ) / \Gamma
( t \rightarrow W^+ b )
\end{eqnarray}
where $\Gamma ( t \rightarrow W^+ b )$ is the partial width for $t
\rightarrow W^+ b$.

\begin{center}
{\bf Acknowledgments}
\end{center}

One of us (CAN) thanks top-quark experimentalists and theorists
for discussions. This work was partially supported by U.S. Dept.
of Energy Contract No. DE-FG 02-86ER40291.

\begin{appendix}

\section{Appendix:  Kinematic Formulas}

In the $(t\overline{t})_{c.m.}$ frame, the angles $\theta _{1,2}$ of the $%
W_{1}^{+}$, $W_{2}^{-}$ and their respective energies $E_{1,2}$
are related by
\begin{equation}
2\widetilde{P}p_{W}\cos \theta _{1,2}=2\widetilde{P}%
_{0}E_{1,2}-m_{t}^{2}-m_{W}^{2}
\end{equation}
where $t$-energy and magnitude of $t$-momentum are
 $\widetilde{P}_{0}=\sqrt{s}/2$, $\widetilde{P}=
\sqrt{\widetilde{P}
_{0}^{2}-m_{t}^{2}} $, and $p_{W}^{2}=E_{1,2}^{2}-m_{W}^{2}$.  \ In the $t_{1}$ %
rest frame,  $\overline{t}_{2}$ rest frame, respectively
\begin{equation}
\theta _{1,2}^{t}=\arccos [\frac{-\sqrt{s}%
(m_{t}^{2}+m_{W}^{2})+4E_{1,2}m_{t}^{2}}{(m_{t}^{2}-m_{W}^{2})\sqrt{%
s-4m_{t}^{2}}}],0\leq \theta _{1,2}^{t}\leq \pi
\end{equation}
which give the kinematic limits
\begin{equation}
E_{1,2}^{\max \mathrm{,}\min }=\frac{\sqrt{s}(m_{t}^{2}+m_{W}^{2})}{%
4m_{t}^{2}}\pm \frac{\sqrt{s}(m_{t}^{2}-m_{W}^{2})}{4m_{t}^{2}}[1-\frac{%
4m_{t}^{2}}{s}]^{1/2}
\end{equation}
The angles $\theta _{1,2}$ are determined uniquely from $\cos
\theta _{1,2}$ and $\sin \theta _{1,2}$ of
\begin{eqnarray}
p_{1,2}\cos \theta _{1,2} &=&\gamma (p_{1,2}^{t}\cos \theta
_{1,2}^{t}+\beta
E_{1,2}^{t}) \\
p_{1,2}\sin \theta _{1,2} &=&p_{1,2}^{t}\sin \theta _{1,2}^{t}
\end{eqnarray}
where $p_{1,2}^{t}=(m_{t}^{2}-m_{W}^{2})/2m_{t}$, $E_{1,2}^{t}=\sqrt{%
(p_{1,2}^{t})^{2}+m_{W}^{2}}$, and $\gamma =\sqrt{s}/(2m_{t})$,
$\beta $ are for
the relativistic boosts between the $(t\overline{t})_{c.m.}$ frame and the $%
t_{1}$, $\overline{t}_{2}$ rest frames. \ A check is
$E_{1,2}=\gamma (E_{1,2}^{t}+\beta p_{1,2}^{t}\cos \theta
_{1,2}^{t})$.

From $\theta _{1,2}$ there is a unique relation between $\cos \psi $ and $%
\cos \phi $,
\begin{equation}
\cos \psi =-\cos \theta _{1}\cos \theta _{2}+\sin \theta _{1}\sin
\theta _{2}\cos \phi
\end{equation}
or equivalently from $\theta _{1,2}^{t}$%
\begin{equation}
\sin \theta _{1}^{t}\sin \theta _{2}^{t}\cos \phi =\frac{4m_{t}^{2}}{%
(m_{t}^{2}-m_{W}^{2})^{2}}\left\{
\begin{array}{c}
p_{1}p_{2}\cos \psi  \\
+\frac{(\sqrt{s}E_{1}-m_{t}^{2}-m_{W}^{2})(\sqrt{s}E_{2}-m_{t}^{2}-m_{W}^{2})%
}{s-4m_{t}^{2}}
\end{array}
\right\}
\end{equation}

The sign of the quantity $\sin \phi $ is the same as the sign of
the auxiliary variable $\sin \Phi _{2}$.

\section{Appendix: Translation Between $\Gamma( \lambda_W, {\lambda_W}^{'}
)$'s Notation and Helicity Parameter's of Refs. [5,11,12,14]}

For the $t\rightarrow W^{+}b$ helicity amplitudes,\ in terms of
the helicity-parameters of Refs. [5,11,12,14], the $\lambda
_{b}=-1/2$ polarized-partial-widths and
W-boson-LT-interference-widths are
\begin{eqnarray}
\Gamma (0,0) & \equiv & \frac{\Gamma }{4}\cdot \{1+\xi +\zeta
+\sigma \}
 \\
\Gamma (-1,-1) & \equiv & \frac{\Gamma }{4}\cdot \{1+\xi -\zeta
-\sigma \}
\\
\Gamma _{\mathit{R}}(0,-1) & \equiv & \frac{\Gamma }{2} \cdot \{\eta +\omega \}  =  \, \Gamma \cdot \eta_L  \\
\Gamma _{\mathit{I}}(0,-1) & \equiv & - \frac{\Gamma }{2} \cdot
\{\eta ^{^{\prime }}+\omega ^{^{\prime }}\} =  \, - \Gamma \cdot
{\eta_L}^{'}
\end{eqnarray}
where the $L$ superscript is suppressed, and $\Gamma$ is the partial width for $t\rightarrow W^+ b$. For $\overline{t}%
\rightarrow W^{-}\overline{b}$, the analogous formulas $\lambda _{\overline{b%
}}=1/2$ polarized-partial-widths and
W-boson-LT-interference-widths are obtained by replacing $-1$
$\rightarrow +1$ in the $\Gamma $'s on the left-hand-sides, and
then barring all of the $\Gamma $'s on both sides and also barring
all the helicity parameters.

The important ${\mathcal{R}}$ suppression factor in (18) was
denoted as $S_W$ in these references.

\section{Appendix:  $\Theta _{B} $ , $\Phi _{R}$ to $\theta _{q}$ , $\phi
_{q}$ Formulas}

The transformation formulas to express the beam spherical angles
$\Theta _{B} $ , $\Phi _{R}$ in terms of $\theta _{q}$ , $\phi
_{q}$ involve the $(t\bar{t})_{c.m.}$ W-boson angles $\theta_1$,
$\theta_2$, and also the auxiliary variables $\sin \Phi _{2}$ and
$\cos \Phi _{2}$ of (69-70) [ see Figs. 8-9]. \ In the
helicity-conserving contributions
\begin{eqnarray}
\cos \Theta _{B} &=&{\mathcal{P}}_{1}+{\mathcal{Q}}_{1}   \\
{\mathcal{P}}_{1} &=&\cos \theta _{1}\cos \theta _{q}-\cos \phi
_{q}\sin \theta _{1}\sin \theta _{q}\cos \Phi
_{2},{\mathcal{Q}}_{1}=-\sin \phi _{q}\sin \theta _{1}\sin \theta
_{q}\sin \Phi _{2}\nonumber
\end{eqnarray}
\begin{eqnarray}
(1+\cos ^{2}\Theta _{B}) &=&{\mathcal{P}}_{0}+{\mathcal{Q}}_{0}   \\
{\mathcal{P}}_{0} &=&1+\cos ^{2}\theta _{1}\cos ^{2}\theta _{q}+\frac{1}{2}%
\sin ^{2}\theta _{1}\sin ^{2}\theta _{q}\nonumber   \\
&&-\cos \phi _{q}\sin 2\theta _{q}\cos \theta _{1}\sin \theta
_{1}\cos \Phi _{2}+\frac{1}{2}\cos 2\phi _{q}\sin ^{2}\theta
_{1}\sin ^{2}\theta
_{q}\cos 2\Phi _{2}\nonumber   \\
{\mathcal{Q}}_{0} &=&-\sin \phi _{q}\sin 2\theta _{q}\cos \theta
_{1}\sin \theta _{1}\sin \Phi _{2}+\frac{1}{2}\sin 2\phi _{q}\sin
^{2}\theta _{1}\sin ^{2}\theta _{q}\sin 2\Phi _{2}\nonumber
\end{eqnarray}
\begin{eqnarray}
\sin ^{2}\Theta _{B}\cos (2\Phi _{R}+\phi ) &=&\mathcal{P}_{\kappa }+%
\mathcal{Q}_{\kappa }   \\
\mathcal{P}_{\kappa } &=&\mathcal{C}\cos \phi
+\mathcal{S}^{^{\prime }}\sin \phi ,\,\,\mathcal{Q}_{\kappa
}=\mathcal{S}\cos \phi -\mathcal{C}^{^{\prime }}\sin \phi
\nonumber
\end{eqnarray}
\begin{eqnarray}
\sin ^{2}\Theta _{B}\sin (2\Phi _{R}+\phi )
&=&\mathcal{P}_{^{\kappa \prime
}}+\mathcal{Q}_{^{\kappa \prime }}   \\
\mathcal{P}_{^{\kappa \prime }} &=&\mathcal{C}^{^{\prime }}\cos \phi +%
\mathcal{S}\sin \phi ,\,\,\mathcal{Q}_{^{\kappa \prime }}=-\mathcal{S}%
^{^{\prime }}\cos \phi +\mathcal{C}\sin \phi \nonumber
\end{eqnarray}
where
\begin{eqnarray}
\mathcal{C} &=&\frac{1}{2}\sin ^{2}\theta _{1}(3\cos ^{2}\theta _{q}-1)%
\nonumber   \\
&&+\cos \phi _{q}\sin 2\theta _{q}\cos \theta _{1}\sin \theta
_{1}\cos \Phi _{2}+\frac{1}{2}\cos 2\phi _{q}\sin ^{2}\theta
_{q}[1+\cos ^{2}\theta _{1}]\cos 2\Phi _{2} \nonumber
\end{eqnarray}
\begin{eqnarray}
{\mathcal{S}}=\sin \phi _{q}\sin 2\theta _{q}\cos \theta _{1}\sin
\theta _{1}\sin \Phi _{2}+\frac{1}{2}\sin 2\phi _{q}\sin
^{2}\theta _{q}[1+\cos ^{2}\theta _{1}]\sin 2\Phi _{2} \nonumber
\end{eqnarray}
\begin{eqnarray}
{\mathcal{C}^{^{\prime }}}=\sin \phi _{q}\sin 2\theta _{q}\sin
\theta _{1}\cos \Phi _{2}+\sin 2\phi _{q}\sin ^{2}\theta _{q}\cos
\theta _{1}\cos 2\Phi _{2} \nonumber
\end{eqnarray}
\begin{eqnarray}
{{\mathcal{S}^{^{\prime }}}}=\cos \phi _{q}\sin 2\theta _{q}\sin
\theta _{1}\sin \Phi _{2}+\cos 2\phi _{q}\sin ^{2}\theta _{q}\cos
\theta _{1}\sin 2\Phi _{2} \nonumber
\end{eqnarray}

For the mixed-helicity contributions, we first define functions of
the final angles
\begin{eqnarray}
{\mathcal{C}}_{1}^{m} &=&\sin \phi _{q}\sin \theta _{q}\cos \Phi
_{2},\quad {\mathcal{S}}_{1}^{m}=\cos \phi _{q}\sin \theta
_{q}\sin \Phi _{2}\nonumber \\
{\mathcal{C}}_{2}^{m} &=&\cos \theta _{q}\sin \theta _{1}+\cos
\phi _{q}\sin
\theta _{q}\cos \theta _{1}\cos \Phi _{2}\nonumber   \\
{\mathcal{S}}_{2}^{m} &=&\sin \phi _{q}\sin \theta _{q}\cos \theta
_{1}\sin
\Phi _{2}\nonumber   \\
{\mathcal{C}}_{3}^{m} &=&\frac{1}{2}\sin \phi _{q}\sin 2\theta
_{q}\cos \theta _{1}\cos \Phi _{2}-\frac{1}{2}\sin 2\phi _{q}\sin
^{2}\theta _{q}\sin \theta
_{1}\cos 2\Phi _{2}\nonumber   \\
{\mathcal{S}}_{3}^{m} &=&\frac{1}{2}\cos \phi _{q}\sin 2\theta
_{q}\cos \theta _{1}\sin \Phi _{2}-\frac{1}{2}\cos 2\phi _{q}\sin
^{2}\theta _{q}\sin \theta
_{1}\sin 2\Phi _{2}\nonumber   \\
{\mathcal{C}}_{4}^{m} &=&\frac{1}{4}\sin 2\theta _{1}(3\cos ^{2}\theta _{q}-1)%
\nonumber   \\
&&+\frac{1}{2}\cos \phi _{q}\sin 2\theta _{q}\cos 2\theta _{1}\cos \Phi _{2}-%
\frac{1}{4}\cos 2\phi _{q}\sin ^{2}\theta _{q}\sin 2\theta
_{1}\cos 2\Phi
_{2}\nonumber   \\
{\mathcal{S}}_{4}^{m} &=&\frac{1}{2}\sin \phi _{q}\sin 2\theta
_{q}\cos 2\theta _{1}\sin \Phi _{2}-\frac{1}{4}\sin 2\phi _{q}\sin
^{2}\theta _{q}\sin 2\theta _{1}\sin 2\Phi _{2}
\end{eqnarray}
Using these definitions,
\begin{eqnarray}
\sin \Phi _{R}\sin \Theta _{B} &=&{\mathcal{C}}_{1}^{m}-{\mathcal{S}}%
_{1}^{m},\cos \Phi _{R}\sin \Theta _{B}={\mathcal{C}}_{2}^{m}+{\mathcal{S}}%
_{2}^{m}\nonumber   \\
\sin \Phi _{R}\sin \Theta _{B}\cos \Theta _{B} &=&{\mathcal{C}}_{3}^{m}-%
{\mathcal{S}}_{3}^{m}\nonumber   \\
\cos \Phi _{R}\sin \Theta _{B}\cos \Theta _{B} &=&{\mathcal{C}}_{4}^{m}+%
{\mathcal{S}}_{4}^{m}
\end{eqnarray}
and
\begin{eqnarray}
\sin (\Phi _{R}+\phi )\sin \Theta _{B} &=&{\mathcal{P}}_{1}^{m}+{\mathcal{Q}}%
_{1}^{m}\nonumber   \\
{\mathcal{P}}_{1}^{m} &=&{\mathcal{C}}_{1}^{m}\cos \phi
+{\mathcal{S}}_{2}^{m}\sin
\phi ,\,\,{\mathcal{Q}}_{1}^{m}=-{\mathcal{S}}_{1}^{m}\cos \phi +{\mathcal{C}}%
_{2}^{m}\sin \phi \nonumber   \\
\cos (\Phi _{R}+\phi )\sin \Theta _{B} &=&{\mathcal{P}}_{2}^{m}+{\mathcal{Q}}%
_{2}^{m}\nonumber   \\
{\mathcal{P}}_{2}^{m} &=&{\mathcal{C}}_{2}^{m}\cos \phi
+{\mathcal{S}}_{1}^{m}\sin
\phi ,\,\,{\mathcal{Q}}_{2}^{m}={\mathcal{S}}_{2}^{m}\cos \phi -{\mathcal{C}}%
_{1}^{m}\sin \phi \nonumber   \\
\sin (\Phi _{R}+\phi )\sin \Theta _{B}\cos \Theta _{B} &=&{\mathcal{P}}%
_{3}^{m}+{\mathcal{Q}}_{3}^{m}\nonumber   \\
{\mathcal{P}}_{3}^{m} &=&{\mathcal{C}}_{3}^{m}\cos \phi
+{\mathcal{S}}_{4}^{m}\sin
\phi ,\,\,{\mathcal{Q}}_{3}^{m}=-{\mathcal{S}}_{3}^{m}\cos \phi +{\mathcal{C}}%
_{4}^{m}\sin \phi \nonumber   \\
\cos (\Phi _{R}+\phi )\sin \Theta _{B}\cos \Theta _{B} &=&{\mathcal{P}}%
_{4}^{m}+{\mathcal{Q}}_{4}^{m}\nonumber   \\
{\mathcal{P}}_{4}^{m} &=&{\mathcal{C}}_{4}^{m}\cos \phi
+{\mathcal{S}}_{3}^{m}\sin
\phi ,\,\,{\mathcal{Q}}_{4}^{m}={\mathcal{S}}_{4}^{m}\cos \phi -{\mathcal{C}}%
_{3}^{m}\sin \phi
\end{eqnarray}

For the ``helicity-flip'' contributions, $\sin ^{2}\Theta
_{B}=2-{\mathcal{P} }_{0}-{\mathcal{Q}}_{0}$ .

\section{Appendix:  $ e \bar{e} \rightarrow t \bar{t} $ Production}

In $ e \bar{e} \rightarrow t \bar{t} $ production, as the
center-of-mass energy increases, the helicity-flip amplitudes
$T(\lambda_1, \lambda_2)$ of (56) will be suppressed relative to
the helicity-conserving ones by the factor of $\sqrt{2} m_t/
(\sqrt{s})$.  With respect to more accurate and more precise
measurements, this could be a useful variable-dependence. We
neglect $m_e / \sqrt{s} $ corrections.  For the case of $t
\bar{t}$ production via $\gamma^{*}$, the formulas in the text
apply with the replacement $ g^2\rightarrow \frac{2}{3} e^2 $ with
$e =\sqrt{4\pi \alpha}$. For $Z^{*}$ production, $ \tilde{T}(-+) =
v_e + a_e $ and $ \tilde{T}(+-) = v_e - a_e $ with $ v_e = e (-1 +
4 \sin^2 { \theta_W} ) / ( 4 \sin{\theta_W} \cos{\theta_W} )$ and
$ a_e = - e / ( 4 \sin{\theta_W} \cos{\theta_W} )$, and $ {T}(-+)
= v_t + a_t (2\widetilde{P} /  \sqrt{s} )$, $ {T}(+-) = v_t - a_t
(2\widetilde{P} /  \sqrt{s} )$, $ {T}(++) =  {T}(--) = \sqrt{2}
v_t m_t/ \sqrt{s} )$, with $ v_t = e (3 - 8 \sin^2 { \theta_W} ) /
( 12 \sin{\theta_W} \cos{\theta_W} )$ and $ a_t = e / ( 4
\sin{\theta_W} \cos{\theta_W} )$ with $\widetilde{P}=$ magnitude
of $t$-momentum in $(t\bar{t})_{cm}$, and $1/s \rightarrow
1/(s-{M_Z}^2)$.

\end{appendix}

\newpage

\begin{center}
{\bf Figure Captions}
\end{center}

FIG. 1:  In the $ (t \bar{t} )_{c.m.}$ frame, the ``final
coordinate system" $ ( \hat{x}, \hat{y}, \hat{z} ) $ for
specification of the beam direction by the spherical angles
$\theta_q$, $\phi_q$.  Note that $\psi$ is the smaller angle
between the ${W_1}^+$ and ${W_2}^{-}$ momenta. For the sequential
decay $ t \rightarrow W^{+} b $ followed by $ W^+ \rightarrow l^+
\nu  $, the spherical angles $ \theta_a$, $\phi_a $ specify the $
l^+ $ momentum in the ${W_1}^+$ rest frame when there is first a
boost from the $ (t \bar{t} )_{c.m.}$  frame to the $t_1$ rest
frame, and then a second boost from the $t_1$ rest frame to the
${W_1}^+$ rest frame, see Fig. 5 below.  The $0^o$ direction for
the azimuthal angle $\phi_a$ is defined by the projection of the
${W_2}^-$ momentum direction.

FIG. 2: Supplement to Fig. 1 to specify the $CP$-conjugate
sequential decay $ \bar{t} \rightarrow W^{-}\bar{b} $ followed by
$ W^- \rightarrow l^- \bar{ \nu }$.  The spherical angles $
\theta_b$, $\phi_b $ specify the $ l^- $ momentum in the ${W_2}^-$
rest frame when ${W_1}^+$ rest frame when there is first a boost
from the $ (t \bar{t} )_{c.m.}$ frame to the $\bar{t_2}$ rest
frame, and then a second boost from the $\bar{t_2}$ rest frame to
the ${W_2}^-$ rest frame.  The $0^o$ direction for the azimuthal
angle $\phi_b$ is defined by the projection of the ${W_1}^+$
momentum direction.  To better display other angles, the values of
the angle $\psi$ are different in Figs. 1 and 2.

FIG. 3: Summary illustration showing the three angles $\theta
_1^t$, $\theta _2^t$ and $\phi $ describing the first stage in the
sequential-decays of the $t\bar{t}$ system in which $%
t_1 \rightarrow {W_1}^{+}b$ and $\bar{t_2} \rightarrow
{W_2}^{-}\bar b$. In (a) the $b$ momentum, not shown, is back to
back with the ${W_1}^{+}$.  In (b) the $\bar{b}$ momentum, not
shown, is back to back with the ${W_2}^{-}$.  From (a) a boost
along the negative ${z_1}^t$ axis transforms the kinematics from
the $t_1$ rest frame to the $ (t \bar{t} )_{c.m.}$ frame and, if
boosted further, to the $\bar{t_2}$ rest frame shown in (b). In
this figure, $ \phi_1 $ of Fig. 4 is shown equal to zero for
simplicity of illustration.

FIG. 4: The usual helicity angles ${\theta_1}^t$ and $\phi_1$
specify the ${W_1}^+$ momentum, in the $t_1$ rest frame, with
$\bar{t_2}$ moving in the negative $z$ direction.  The polar angle
${\theta_2}^t$ for the ${W_2}^-$ is defined analogously in the
$\bar{t_2}$ rest frame, c.f. Fig. 3.  The azimuthal angles
$\phi_1$ and $\phi_2$ are Lorentz invariant under boosts along the
${z_1}^t$ axis. The sum $\phi = \phi_1 + \phi_2 $ is the angle
between the $t_1$ and $\bar{t_2}$ decay planes.

FIG. 5: The two pairs of spherical angles ${\theta_1}^t, \phi_1$
and $\theta_a$,$ \tilde{\phi_a}$ specify the two stages in the
sequential decay $ t \rightarrow W^{+} b \rightarrow (l^+ \nu ) b
$ when the boost to the ${W_1}^+$ rest frame is from the $t_1$
rest frame.  In the ${W_1}^+$ rest frame, to reference the $0^o$
direction for $\tilde{\phi_a}$ the axis $x_a$ lies in the
$\bar{t_2}$ half-plane.  In this figure, $\phi_1$ of Fig. 4 is
shown equal to zero for simplicity of illustration. Similarly, two
pairs of spherical angles ${\theta_2}^t, \phi_2$ and $\theta_b$,$
\tilde{\phi_b}$ specify the two stages in the $CP$-conjugate
sequential decay $ \bar{t} \rightarrow W^{-} \bar{b} $ followed by
$ W^- \rightarrow l^- \bar{ \nu} $ when the boost is from the
$\bar{t_2}$ rest frame.

FIG. 6: The derivation of the general ``beam referenced
stage-two-spin-correlation" function begins in the ``home" or
starting coordinate system $ ( {x_h}, {y_h}, {z_h} ) $ in the $ (t
\bar{t} )_{c.m.}$ frame.  $t_1$ is moving in the positive $z_h$
direction, and $\theta_1, \phi_1$ specify the ${W_1}^+$ momentum
direction.  The beam direction is specified by the spherical
angles $\Theta_B, \Phi_B$.  Note that $\Phi_R = \Phi_B - \phi_1$.

FIG. 7:  Supplement to previous figure to show $\theta_2, \phi_2$
which specify the ${W_2}^-$ momentum direction.

FIG. 8: In the derivation, the ``barred" coordinate system $ (
\bar{x}, \bar{y}, \bar{z} ) $ in the $ (t \bar{t} )_{c.m.}$ frame
has ${W_1}^+$ along the positive $\bar{z}$ axis with the $t_1$ in
the negative $\bar{x}$ half-plane.  A rotation by $\theta_1$ has
transformed the description from the previous ``home system" to
the one in this ``barred" coordinate system.

FIG. 9: Supplement to previous figure, to show specification of
the ${W_2}^-$ by the spherical angles $ \Theta_2, \Phi_2$. Note
that $ \psi +  \Theta_2 = \pi $.  A further rotation by minus
$\Phi_2$ about the $\bar{z}$ axis transforms this ``barred system"
 description" into that in the ``final coordinate system" shown
in Figs. 1 and 2.

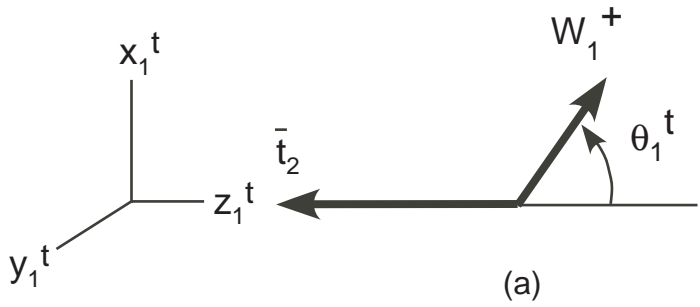
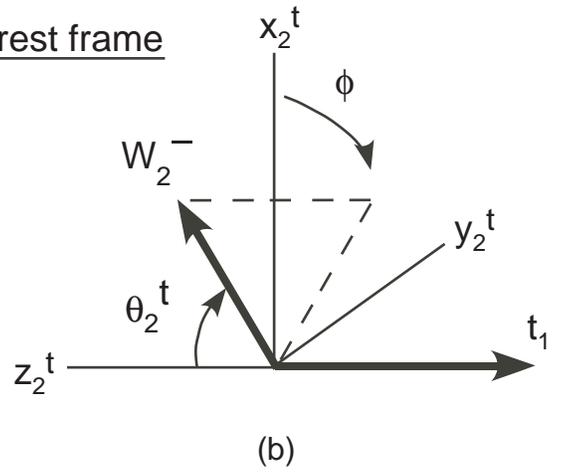
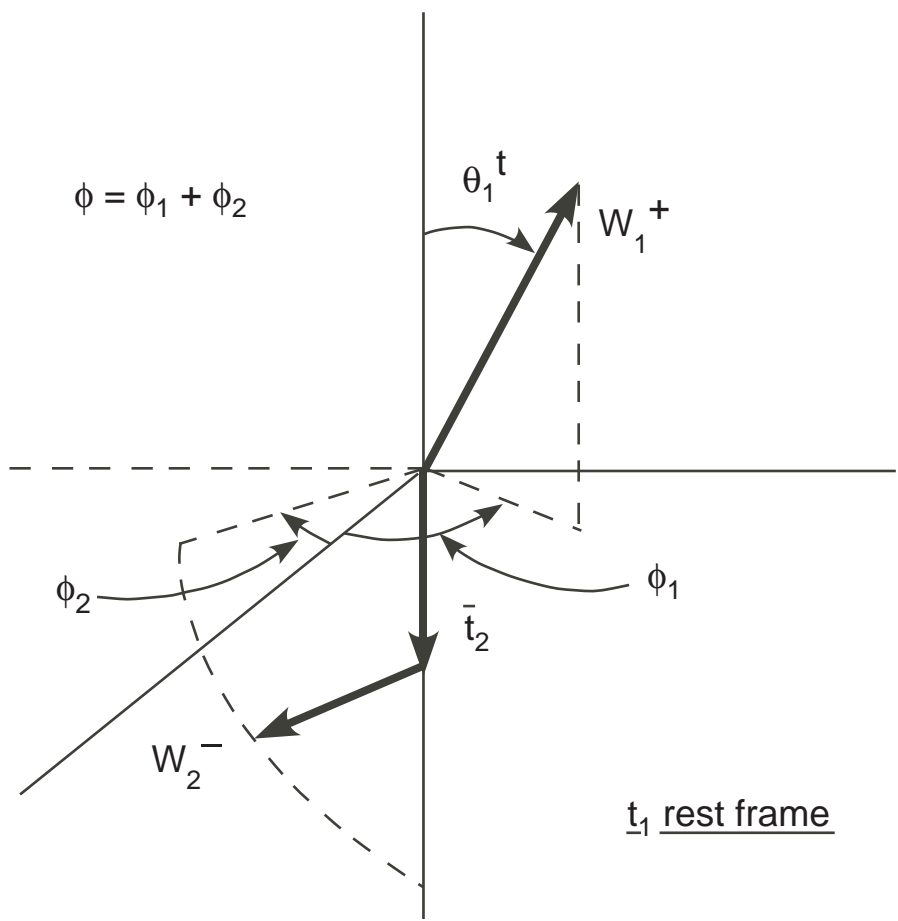
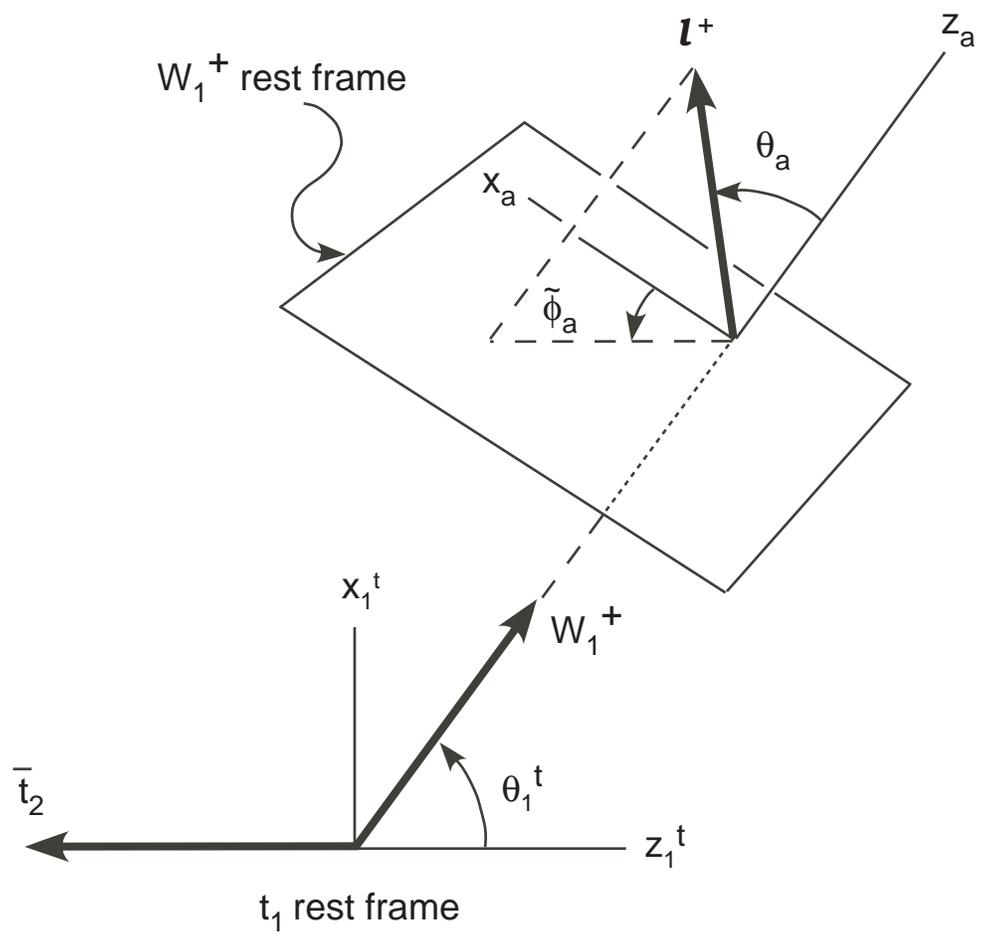
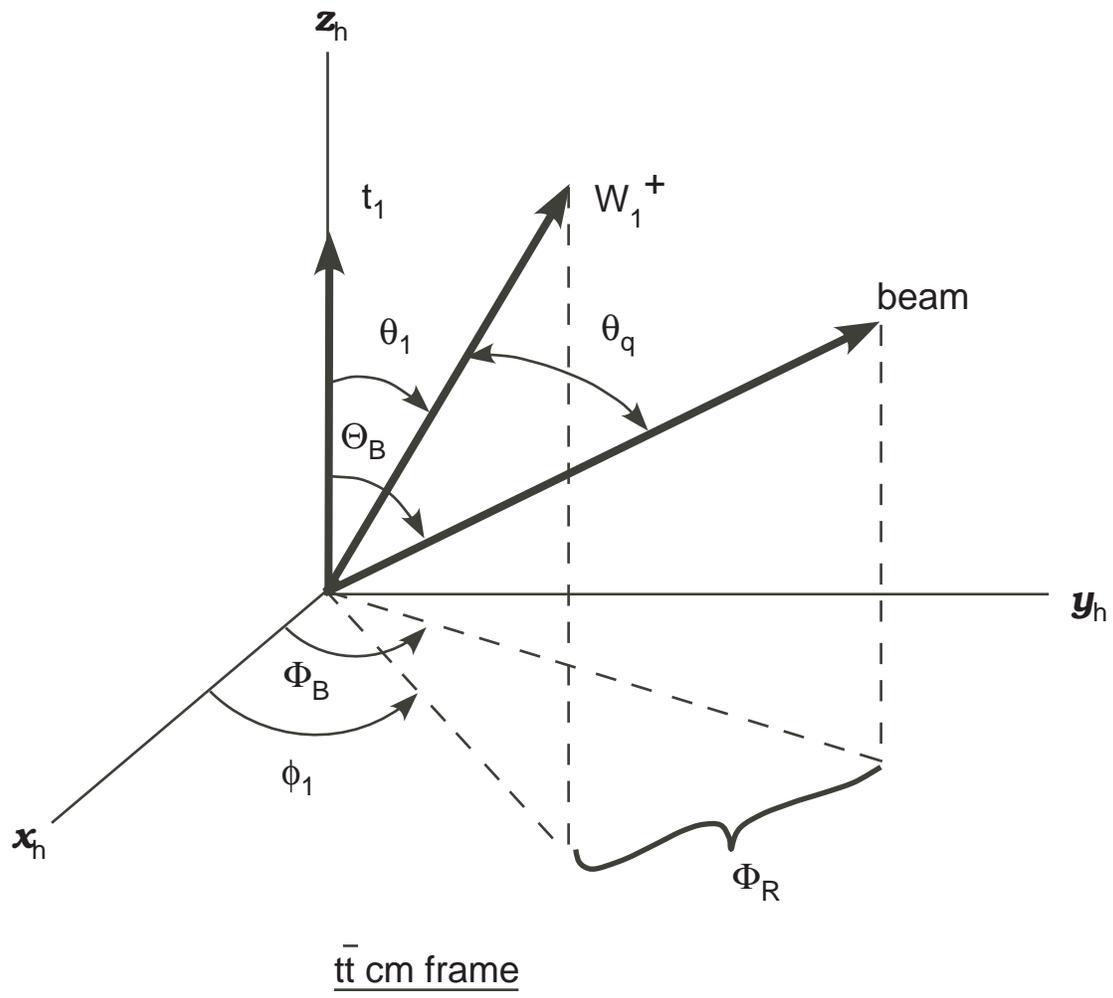
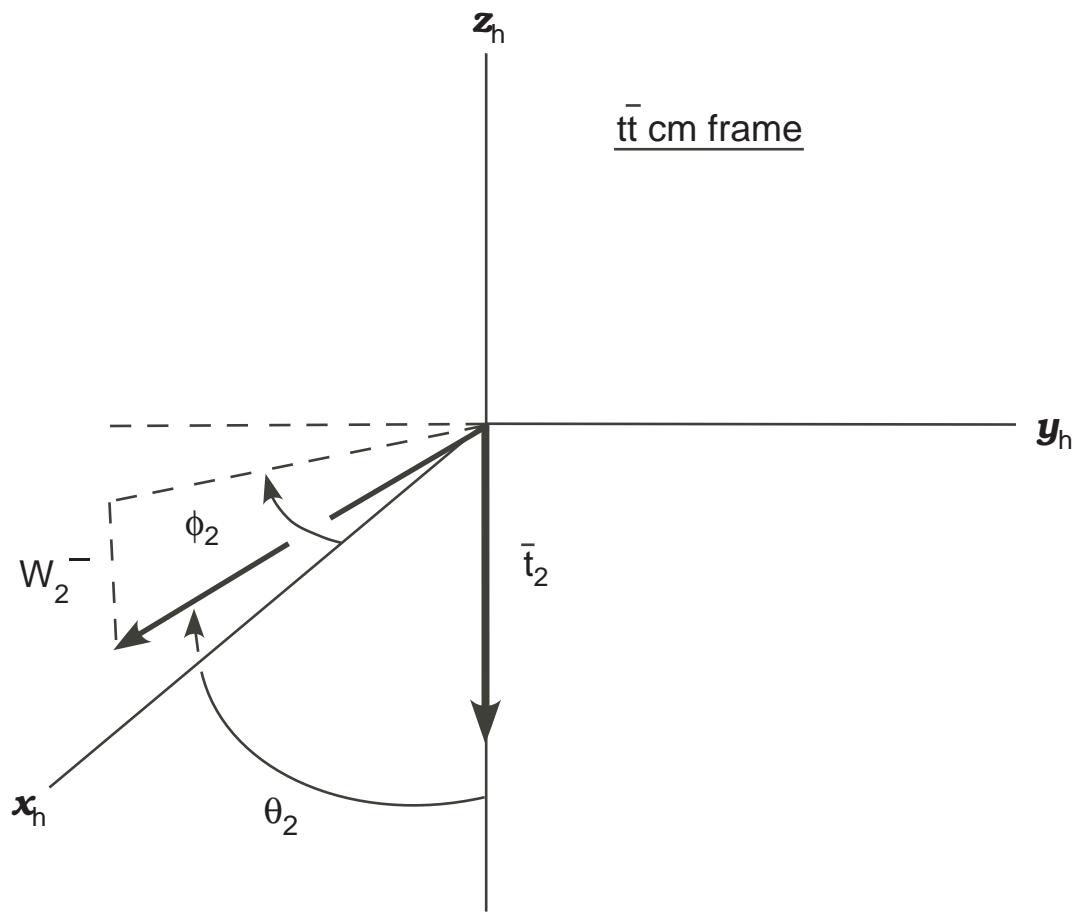
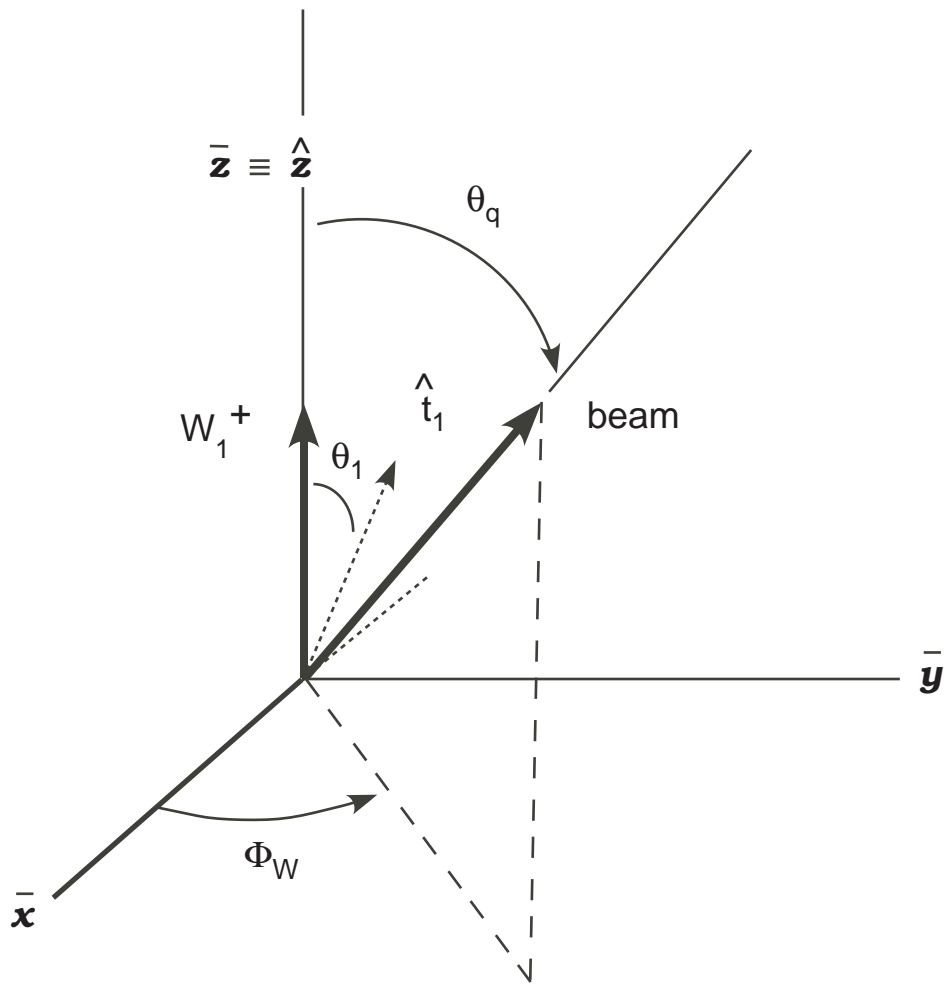
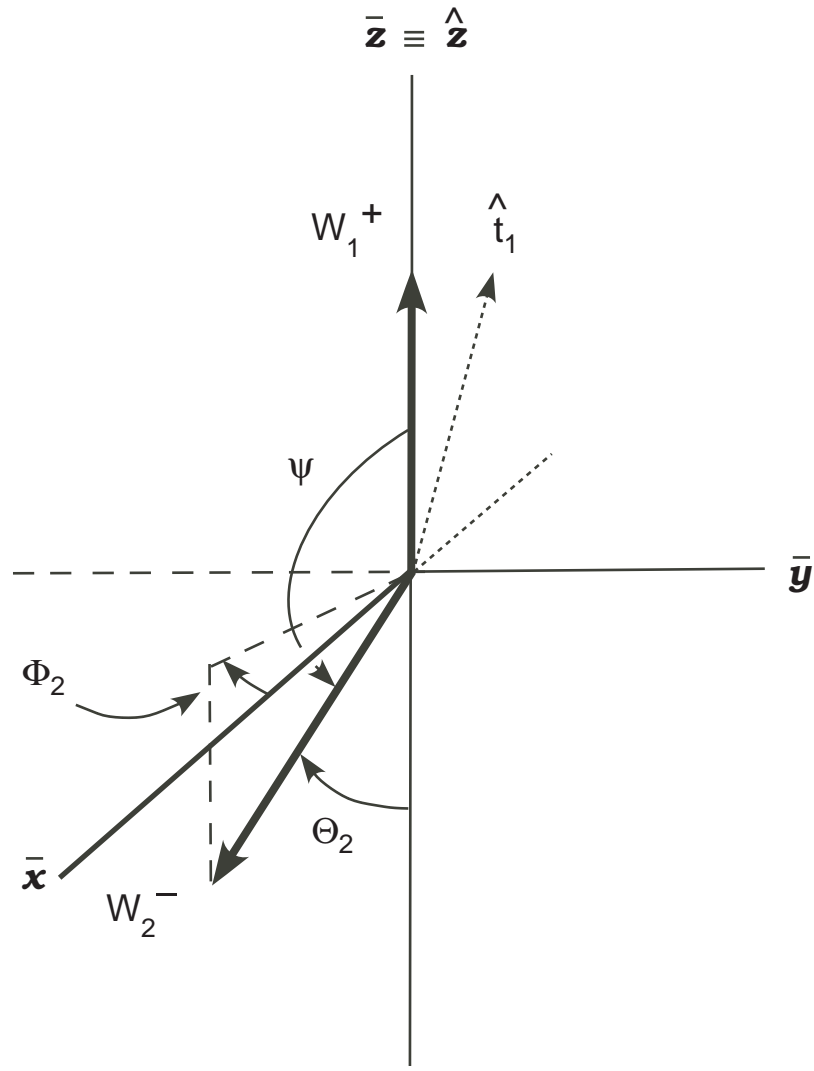
\end{document}